\documentclass[conference]{IEEEtran}

\usepackage{cite}
\usepackage{amsmath,amssymb,amsfonts}
\usepackage{algorithmic}
\usepackage{graphicx}
\usepackage{textcomp}
\usepackage{xcolor}
\def\BibTeX{{\rm B\kern-.05em{\sc i\kern-.025em b}\kern-.08em
    T\kern-.1667em\lower.7ex\hbox{E}\kern-.125emX}}

\usepackage{amssymb}
\usepackage{bm, bbm}
\usepackage{mathtools}
\usepackage{commath}
\usepackage{stmaryrd}
\usepackage{faktor}

\usepackage{graphicx, wrapfig}
\usepackage{forest}
\usepackage{tikz-cd}
\usepackage[position=bottom]{subfig}

\usepackage{array,multirow,booktabs}
\usepackage{makecell} 
\usepackage{tabularx} 
\usepackage{rotating} 

\usepackage[shortlabels]{enumitem}

\newcommand{\given}{\vert}
\newcommand{\mgiven}{\;\middle\vert\;}

\renewcommand{\hat}{\widehat}
\renewcommand{\tilde}{\widetilde}
\renewcommand{\bar}{\overline}

\renewcommand{\d}{\mathrm{d}}

\newcommand{\error}{\mathrm{error}}

\renewcommand{\r}{\mathrm{r}}
\renewcommand{\sp}{\mathrm{sp}}
\DeclareMathOperator{\dotle}{\,\dot{\le}\,}
\DeclareMathOperator{\dotge}{\,\dot{\ge}\,}

\newcommand{\1}{\mathbbm{1}}

\newcommand{\E}{\mathbb{E}}
\newcommand{\N}{\mathbb{N}}

\newcommand{\R}{\mathbb{R}}

\newcommand{\Pbb}{\mathbb{P}}

\newcommand{\Ccal}{\mathcal{C}}

\newcommand{\Pcal}{\mathcal{P}}

\newcommand{\Scal}{\mathcal{S}}

\newcommand{\Tcal}{\mathcal{T}}
\newcommand{\Vcal}{\mathcal{V}}

\newcommand{\Xcal}{\mathcal{X}}
\newcommand{\Ycal}{\mathcal{Y}}

\newcommand{\xv}{\pmb{x}}
\newcommand{\yv}{\pmb{y}}

\newcommand{\Xv}{\pmb{X}}
\newcommand{\Yv}{\pmb{Y}}

\renewcommand{\Im}{\mathrm{Im}}

\newcommand{\Bin}{\mathsf{Bin}}

\newtheorem{theorem}{Theorem}
\newtheorem{lemma}[theorem]{Lemma}
\newtheorem{proposition}[theorem]{Proposition}
\newtheorem{corollary}[theorem]{Corollary}

\newtheorem{remark}{Remark}

\usepackage{color}
\colorlet{RED}{red}

\newcommand{\bL}{L}
\allowdisplaybreaks

\begin{document}

\title{Error Exponents for Randomised List Decoding}

 \author{%
	\IEEEauthorblockN{Henrique~K.~Miyamoto and Sheng~Yang}
	\IEEEauthorblockA{
		Laboratoire des Signaux et Systèmes (L2S)\\
		Université Paris-Saclay, CNRS, CentraleSupélec\\
		Gif-sur-Yvette, France\\
		Email: \{henrique.miyamoto, sheng.yang\}@centralesupelec.fr}
}

\maketitle

\begin{abstract}
	This paper studies random-coding error exponents of randomised list decoding, in which the decoder randomly selects $L$ messages with probabilities proportional to the decoding metric of the codewords.
	The exponents (or bounds) are given for mismatched, and then particularised to matched and universal decoding metrics.
	Two regimes are studied: for fixed list size, we derive an ensemble-tight random-coding error exponent, and show that, for the matched metric, it does not improve the error exponent of ordinary decoding.
	For list sizes growing exponentially with the block-length, we provide a non-trivial lower bound to the error exponent that is tight at high rates under the matched metric.
\end{abstract}

\section{Introduction}

In list decoding, instead of outputting one estimate of the message sent by the transmitter, the decoder produces a list of $L$ candidates. In this case, a list-decoding error is said to occur if the correct message is absent from that list. The case ${L=1}$ recovers the ordinary decoding problem. In this work, we are interested in information-theoretic aspects of list decoding, specifically, random-coding error exponents \cite{elias1957,wozencraft1958,shannon1967,forney1968,dyachkov1980,blinovsky2001,hof2010,merhav2014,somekh-baruch2019,bondaschi2022}.

We consider that the size~$L$ of the list is set before transmission occurs, and does not depend on the channel realisation---as in the initial works~\cite{elias1957,wozencraft1958}, and differently from, e.g., \cite{forney1968}. Following~\cite{merhav2014}, we distinguish two regimes of interest: in the fixed list size regime, the size of the list is constant and does not depend on the block-length~$n$; in the exponential list size regime, it grows exponentially with the block-length, namely, as $L(n) = e^{n\lambda}$, for some $\lambda>0$.

In \emph{deterministic list decoding}, the list produced by the decoder contains the $L$ messages corresponding to the codewords with the $L$ highest decoding metrics. This setup has been well investigated: a lower bound to the error probability was given in~\cite{shannon1967}; the error exponent under the fixed list size regime was studied in \cite{gallager1968,dyachkov1980,merhav2014}, and under the exponential list size regime in \cite{csiszar2011,merhav2014}. These results quantify how much the error exponents can be improved as function of the list size.

On the other hand, in randomised decoding \cite{yassaee2013,scarlett2015,merhav2017,liu2017,bhatt2018,miyamoto2025} (also known as \emph{stochastic likelihood decoding}, or simply \emph{likelihood decoding}), the decoder selects a (single) message at random, with probabilities that are proportional to the decoding metric of each codeword.
In this case, decoding can be done by sampling codewords~\cite{bhatt2018,miyamoto2025}, instead of solving a discrete optimisation problem (i.e., maximising the decoding metric). The error exponent of both matched and mismatched randomised decoding has been studied in~\cite{scarlett2015,merhav2017}. In particular, it is shown that, under the matched metric, the error exponent of randomised decoding is the same as that of deterministic decoding (see also~\cite{liu2017}).

In this paper, we introduce and investigate \emph{randomised list decoding}, in which the decoder draws randomly (and independently) a list of $L$ messages, proportionally to the decoding metric of the codewords. This can be seen as a natural generalisation of both deterministic list decoding and (ordinary) randomised decoding. Contrary to deterministic decoding, in this case, the list may contain repeated messages, potentially harming the efficiency of the list decoding effect. We ask if this sub-optimality can affect the error exponent, and by how much; or if the same gains of list decoding in the deterministic case can be sustained with the randomised strategy.

Our contributions are as follows. First, in the fixed list size regime, we derive a random-coding error exponent for mismatched metrics that is tight with respect to the ensemble of random codes, and then particularise it for the matched metric. We show that, in the latter case, the error exponent cannot improve over that of ordinary decoding ($L=1$). Our proof is based on extending a technique of~\cite{merhav2017} for $L=1$ to a fixed list size $L\ge1$. Second, in the exponential list size regime, our analysis provide a lower bound to the error exponent for mismatched metrics, via a different proof technique that leverages the result for $L=1$. Specialising to the case of matched metric, we show that the error exponent is tight for rates above a certain critical rate. Third, we show, in both regimes, that the results for matched metrics can be equally obtained with a universal metric that does not depend on the channel law, and is the randomised analogue of the maximum mutual information~(MMI) decoder~\cite{csiszar2011,merhav2017}.

The problem is formalised in Section~\ref{sec:preliminaries}. The results for fixed list size are presented in Section~\ref{sec:fixed-list-size}, and for exponential list size in Section~\ref{sec:exponential-list-size}.

\section{Preliminaries} \label{sec:preliminaries}

\subsection{Notation}

We denote $\xv \coloneqq x_1^n \in \Xcal^n$ a length-$n$ sequence over finite alphabet~$\Xcal$, and $\hat{P}_{\xv}$ its type (empirical distribution) \cite[Ch.~2]{csiszar2011}. We let $\mathcal{P}(\mathcal{X})$ denote the set of distributions over $\mathcal{X}$, $\mathcal{P}_n(\mathcal{X})$ the set of $n$-length types over~$\mathcal{X}$, and $\mathcal{T}_n(P_X)$ the type class of the type $P_X \in \mathcal{P}_n(\mathcal{X})$. Analogous notations apply to joint distributions and types. We add a subscript to information quantities to indicate the underlying distributions, e.g., entropy $H_P(X)$, mutual information $I_P(X \colon Y)$, for $(X,Y) \sim P_{XY}$. For two positive sequences $(a_n)_{n\in\N}$ and $(b_n)_{n\in\N}$, we denote $a_n \doteq b_n$, if $\lim_{n\to\infty} \frac{1}{n}\log \frac{a_n}{b_n} = 0$, and $a_n \dotle b_n$ (or $b_n \dotge a_n$), if $\limsup_{n\to\infty} \frac{1}{n}\log \frac{a_n}{b_n} \le 0$.

\subsection{Problem Setup}

Consider a discrete memoryless channel~(DMC) over input alphabet $\Xcal$ and output alphabet $\Ycal$, with transition probabilities $W(y \given x)$. The channel law for input sequence $\xv \coloneqq x_1^n \in \Xcal^n$ and output sequence $\yv \coloneqq y_1^n \in \Ycal^n$ is $W^n(\yv \given \xv) \coloneqq \prod_{i=1}^{n} W(y_i \given x_i)$. The transmitter employs a code $\Ccal \coloneqq \left( \xv_1, \dots, \xv_M \right) \subseteq \Xcal^n$ of block-length~$n$ and rate $R = (\log M)/n$. It chooses a message $m \in \left\{ 1, \dots, M \right\}$ uniformly at random, and encodes it as codeword $\xv_m \in \Ccal$.

Given a received sequence $\yv \in \Ycal^n$, an ordinary \emph{randomised decoder} selects a message according to the distribution
\begin{equation} \label{eq:decoding-probability-m}
	P_{M \given \Yv}\left( m \,\given\, \yv \right)
	= \frac{u_n(\xv_m,\yv)}{\sum_{m'=1}^{M} u_n(\xv_{m'},\yv)},
\end{equation}
where $u_n \colon \Xcal^n \times \Ycal^n \to \R_+$ is a function called \emph{decoding metric}. We will be interested in metrics that only depend on the sequences $\xv,\yv$ through their joint type $\hat{P}_{\xv,\yv}$. These will be written in the form $u_n(\xv,\yv) = e^{ng(\hat{P}_{\xv,\yv})}$, for some function $g \colon \Pcal(\Xcal \times \Ycal) \to \R$. For instance, in the case of the \emph{matched decoding metric}, we have $u_n(\xv,\yv) = W^n(\yv \given \xv) = e^{ng(\hat{P}_{\xv,\yv})}$, with $g(P_{XY}) = \E_{P}\left[ \log W(Y \given X) \right]$. Another case of interest will be the \emph{MMI metric} $u_n(\xv,\yv) = e^{ng(\hat{P}_{\xv,\yv})}$ with $g(P_{XY}) = I_P(X \colon Y)$ that corresponds to the randomised version of the maximum mutual information~(MMI) decoder~\cite{csiszar2011,merhav2017}.

While an ordinary randomised decoder only outputs one message, a \emph{randomised list decoder} of list size $L$ produces a list of messages $\left( m_1, \dots, m_L \right)$, each of which is independently drawn according to~\eqref{eq:decoding-probability-m}. In contrast to deterministic list decoding, this strategy allows the list to contain repeated messages. Both the fixed and exponential list size regimes are considered.

We are interested in random-coding error exponents, and focus on constant-composition random coding, i.e., the codewords are drawn independently and uniformly among a type class. For a chosen distribution $Q \in \Pcal(\Xcal)$, the random-coding distribution is
$P_{\Xv}(\xv) = \frac{1}{\left| \Tcal_n(Q_n) \right|} \1_{\Tcal_n(Q_n)}(\xv)$ where $Q_n \in \Pcal_n(\Xcal)$ is a type with same support as $Q$ and satisfying $\max_{x\in\Xcal}\left| Q_n(x) - Q(x) \right|_{\infty} \le \frac{1}{n}$. We denote $\overline{p}_{e}(n,L)$ the average error probability (over random codes and channel realisations) for block-length $n$ and list size $L$.
The random-coding error exponent is then
\begin{equation}
	\lim_{n\to\infty} -\frac{1}{n}\log\bar{p}_{e}(n,L).
\end{equation}

Since the messages are uniformly chosen, it is without loss of generality to consider that message $m=1$ was sent. The probability of list-decoding error is the probability that none of the $L$~messages in the list is correct, so it becomes
\begin{align} \label{eq:list-decoding-error-probability}
	\bar{p}_{e}(n,L)
	&= \E\left[ \left( 1 - \frac{u_n(\Xv_1,\Yv)}{\sum_{m'=1}^{M} u_n(\Xv_{m'},\Yv)} \right)^L \right],
\end{align}
where the expectation is taken with respect to
\begin{equation*}
	(\Xv_1,\dots,\Xv_M,\Yv) \sim P_{\Xv}(\xv_1) \cdots P_{\Xv}(\xv_M) W^n(\yv \given \xv_1).
\end{equation*}
A trivial upper bound is $\bar{p}_{e}(n,L) \le \bar{p}_{e}(n,1)$, which corresponds to the average probability of error of ordinary randomised decoding. Another upper bound is the average probability of error of deterministic list decoding with matched metric, which is the optimal list-decoding strategy~\cite{merhav2014}.

\subsection{Classical Error Exponents} \label{subsec:classical-error-exponents}
We review some results and introduce some notation.
The error exponent using any specific code under matched ordinary decoding (either deterministic~\cite{csiszar2011} or randomised~\cite{scarlett2015}) is lower bounded by the \emph{random-coding error exponent}
\begin{align} \label{eq:random-coding-error-exponent}
	E_{\r}(R,Q) \coloneqq \min_{P_{XY} \colon P_X=Q}
	&D\left( P_{XY} \| Q \times W \right) \nonumber \\
	&\quad + \left[ I_P(X \colon Y) - R \right]_+,
\end{align}
and upper bounded by the \emph{sphere packing error exponent}
\begin{equation} \label{eq:sphere-packingerror-exponent}
	E_{\sp}(R,Q) \coloneqq \min_{P_{XY} \colon P_X=Q,\ I_{P}(X \colon Y) \le R} D\left( P_{XY} \| Q \times W \right).
\end{equation}
We recall that \eqref{eq:random-coding-error-exponent} and \eqref{eq:sphere-packingerror-exponent} coincide for rates~$R$ above a certain critical rate~$R_0 \coloneqq R_0(Q)$, which is the smallest rate such that the curve $R \mapsto E_{\sp}(R,Q)$ meets its supporting line of slope $-1$, cf.~\cite[Cor.~10.4]{csiszar2011}.

One of the results of Shannon, Gallager and Berlekamp in~\cite{shannon1967} is that the error exponent of list decoding with list size $L(n) = e^{n\lambda}$ is upper bounded by $E_{\sp}\left( R - \lambda \right)$.

\section{Fixed List Size} \label{sec:fixed-list-size}

\subsection{Main Result and Discussion}

Our main result in the fixed list size regime is an ensemble-tight error exponent.

\begin{theorem}[Mismatched metric] \label{thm:error-exponent-fixed-L-mismatched}
	Under constant-composition random coding with distribution~$Q$, the average error probability of randomised list decoding with decoding metric $u_n(\xv,\yv) = e^{ng(\hat{P}_{\xv,\yv})}$ and fixed list size $L$ satisfies $\lim_{n\to\infty}-\frac{1}{n}\log \bar{p}_{e}(n,L) = E_1(R,Q,L)$, where
	\begin{align} \label{eq:error-exponent-fixed-L-mismatched}
		&E_{1}(R,Q,L) \coloneqq
		\min_{P_{XY} \colon P_X=Q}
		\min_{\tilde{P}_{XY} \colon \tilde{P}_X=Q, \tilde{P}_Y=P_Y} \nonumber\\
		&\hspace{1.5em} D(P_{XY}\| Q \times W)
		+
		\left[ I_{\tilde{P}}(X \colon Y) - R \right]_+ \nonumber\\
		&\hspace{1.5em} + L \left[ g(P_{XY}) - g(\tilde{P}_{XY}) - \left[ R - I_{\tilde{P}}(X \colon Y) \right]_+ \right]_+.
	\end{align}
\end{theorem}

Note that, as a mismatched error exponent, \eqref{eq:error-exponent-fixed-L-mismatched} recovers the results of \cite[Eq.~(16)]{scarlett2015} and \cite[Eq.~(14)]{merhav2017} by setting $L=1$. We now specialise the result to the case of matched and MMI metrics.

\begin{corollary}[Matched metric] \label{cor:error-exponent-fixed-L-matched}
	In the case of the matched metric $g(P_{XY}) = \E_{P} \left[ \log W(Y \given X) \right]$, the error exponent \eqref{eq:error-exponent-fixed-L-mismatched} becomes
	\begin{align} \label{eq:error-exponent-fixed-L-matched}
		&E_1(R,Q,L) = E_{\r}(R,Q) = \nonumber\\
		&\min_{P_{XY} \colon P_X=Q}
		D(P_{XY} \| Q \times W) + \left[ I_{P}(X \colon Y) - R \right]_+.
	\end{align}
	Moreover, the same result is obtained with the MMI metric $g(P_{XY}) = I_P(X \colon  Y)$.
\end{corollary}
\begin{IEEEproof}
	On the one hand, for any metric, we have $E_1(R,Q,L) \ge E_1(R,Q,1)$. In both the cases of matched metric~\cite{scarlett2015} and MMI metric~\cite{merhav2017}, $E_1(R,Q,1) = E_{\r}(R,Q)$, showing that $E_1(R,Q,L) \ge E_{\r}(R,Q)$ in these cases.
	On the other hand, for any metric, choosing $\tilde{P}_{XY} = P_{XY}$ in the inner minimisation of~\eqref{eq:error-exponent-fixed-L-mismatched} yields an upper bound for that error exponent, namely, $E_1(R,Q,L) \le E_{\r}(R,Q)$.
\end{IEEEproof}

\begin{figure}[!t]
	\centering
	\includegraphics[width=0.8\linewidth]{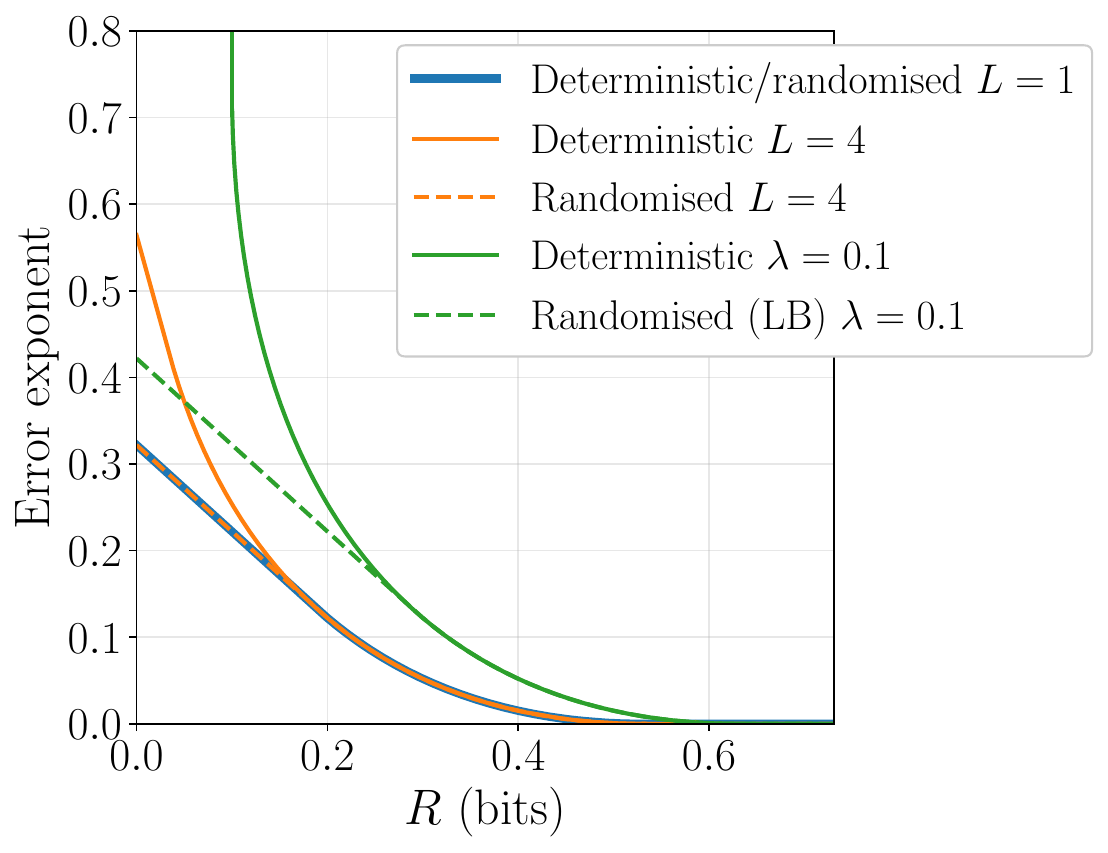}
	\caption{Error exponents in a BSC with cross-over probability~$0.1$ and constant-composition random coding $Q=\left( \frac{1}{2},\frac{1}{2} \right)$, for ordinary decoding (blue), fixed list size~$L=4$ (orange), and exponential list size $L(n)=2^{n\lambda}$ (green). All quantities are in base~$2$. (LB = lower bound.)} 
	\label{fig:example}
\end{figure}

This result is interesting because it shows that \emph{randomised} list decoding with fixed list size $L$ cannot improve the error exponent with respect to ordinary randomised decoding ($L=1$), using the matched decoding metric. This is in contrast with the fixed list size regime in \emph{deterministic} list decoding~\cite{dyachkov1980,merhav2017}, in which the error exponent is
\begin{align} \label{eq:error-exponent-fixed-L-matched-deterministic}
	&\hspace{-0.3em}\tilde{E}_1(R,Q,L) \coloneqq \nonumber\\
	&\hspace{0.2em}\min_{P_{XY} \colon P_X=Q} D(P_{XY} \| Q \times W) + L \left[ I_{P}(X \colon Y) - R \right]_+,
\end{align}
which is the modified random coding exponent of~\cite[Eq.~(10.28)]{csiszar2011}. Intuitively, the reason for this difference, as mentioned, is that randomised list decoding can produce a list with repeated messages, while the deterministic counterpart ensures that the list has $L$ distinct elements, thus effectively reducing the probability of list-decoding error. Nevertheless, \eqref{eq:error-exponent-fixed-L-matched} and \eqref{eq:error-exponent-fixed-L-matched-deterministic} coincide for high enough rates, namely, above the critical rate $R_0$~\cite[Ch.~10]{csiszar2011}. In fact, in that region, list decoding with fixed list size cannot improve the error exponent of ordinary decoding, not even the deterministic version. All these phenomena are illustrated in Fig.~\ref{fig:example} for $L=4$ (orange curves) over a binary symmetric channel~(BSC).

\subsection{Proof (Sketch) of Theorem~\ref{thm:error-exponent-fixed-L-mismatched}}

The proof of this result is an extension of the technique of~\cite{merhav2017} (see also~\cite{merhav2025}) to fixed list size $L \ge 1$, and uses the type class enumeration method~\cite[Ch.~4]{merhav2025}. Here we present a concise version; a detailed version can be found in the Appendix. The probability of error~\eqref{eq:list-decoding-error-probability} can be written as
\begin{align} \label{eq:prob-error-L-0}
	\bar{p}_e(n,L)
	= \sum_{\xv\in\Xcal^n} \sum_{\yv\in\Ycal^n} P_{\Xv}(\xv) W^n(\yv \given \xv) \Pr\left( \error \given \xv,\yv \right),
\end{align}
with
\begin{align*}
	&\Pr\left( \error \given \xv,\yv \right)\\
	&\hspace{1em}= \E\left[ \left( \frac{\sum_{m'\neq1} u_n(\Xv_{m'},\yv)}{u_n(\xv,\yv) + \sum_{m'\neq1} u_n(\Xv_{m'},\yv)} \right)^L \right]\\
	&\hspace{1em}= \E\left[ \left( \frac{\sum_{\tilde{P}_{XY}} N_n(\tilde{P}_{XY}) e^{ng(\tilde{P}_{XY})}}{e^{ng(\hat{P}_{\xv,\yv})} + \sum_{\tilde{P}_{XY}} N_n(\tilde{P}_{XY}) e^{ng(\tilde{P}_{XY})}} \right)^L \right],
\end{align*}
where $N_n(\tilde{P}_{XY}) \coloneqq \sum_{m'=2}^{M} \1_{\Tcal_n(\tilde{P}_{XY})}(\Xv_{m'},\yv)$, and the sums are over types $\tilde{P}_{XY}$ such that $\tilde{P}_X = Q_n$ and $\tilde{P}_Y = \hat{P}_{\yv}$. Using the integral representation $\E[X] = \int_0^1 \Pbb\left( X\ge x \right)\d x$ for the expectation  of a random variable $X \in \left[0,1\right]$, the change of variables $e^{-nL\theta} = t$, and simple algebraic manipulation, we find
\begin{align*}
	&\Pr\left( \error \given \xv,\yv \right) \nonumber\\
	&\hspace{0.5em}= \int_0^1 \Pbb\left( \frac{
		\sum_{\tilde{P}_{XY}} N_n(\tilde{P}_{XY}) e^{ng(\tilde{P}_{XY})}
	}{
		e^{ng(\hat{P}_{\xv,\yv})}
		+ \sum_{\tilde{P}_{XY}} N_n(\tilde{P}_{XY}) e^{ng(\tilde{P}_{XY})}
	}  \ge t^{\frac{1}{L}} \right) \d t \nonumber\\
	&\hspace{0.5em}= nL \int_0^\infty e^{-nL\theta}\\
	&\hspace{2em} \times\Pbb \left(
	\sum_{\tilde{P}_{XY}} N(\tilde{P}_{XY}) e^{ng(\tilde{P}_{XY})}
	\ge \frac{e^{-n\theta}}{1-e^{-n\theta}} e^{ng(\hat{P}_{\xv,\yv})} \right) \d \theta \nonumber.
\end{align*}

Next, denote $\zeta_{\theta}(n) \coloneqq -\frac{1}{n}\log\left( 1-e^{-n\theta} \right) > 0$ and $\delta(n) \coloneqq \frac{|\Xcal||\Ycal|}{n}\log(n+1)$. Recall that the number of types $\tilde{P}_{XY}$ over which the sum acts is upper bounded by $e^{n\delta(n)}$. Then, using well-known properties of the types~\cite[Ch.~4]{merhav2025}, we have
\begin{align}
	&\Pr\left( \error \mgiven \xv,\yv \right) \nonumber\\
	&\ge nL \int_{0}^{\infty} e^{-nL\theta} \nonumber\\
	&\hspace{1em} \times \max_{\tilde{P}_{XY}} \Pbb\left( N_n(\tilde{P}_{XY}) e^{ng(\tilde{P}_{XY})} \ge e^{-n\left(\theta - \zeta_{\theta}(n) - g(\hat{P}_{\xv,\yv}) \right)} \right)\d \theta \nonumber\\
	&\ge nL \int_{0}^{\infty} e^{-nL\theta} e^{-n\delta(n)} \nonumber\\
	&\hspace{1em}\times \sum_{\tilde{P}_{XY}} \Pbb\left( N_n(\tilde{P}_{XY}) e^{ng(\tilde{P}_{XY})} \ge e^{-n\left(\theta - \zeta_{\theta}(n) - g(\hat{P}_{\xv,\yv}) \right)} \right)\d \theta \nonumber\\
	&= nL e^{-n\delta(n)} \sum_{\tilde{P}_{XY}} \int_{0}^{\infty} e^{-nL\theta} \nonumber\\
	&\hspace{3em} \times  \Pbb\bigg( N_n(\tilde{P}_{XY}) e^{ng(\tilde{P}_{XY})} \ge e^{-n\left(\theta - \zeta_{\theta}(n) - g(\hat{P}_{\xv,\yv}) \right)} \bigg)\d \theta \nonumber\\
	&\ge nL e^{-n\delta(n)} \max_{\tilde{P}_{XY}} \int_{0}^{\infty} e^{-nL\theta} \cdot \Pbb\bigg( N_n(\tilde{P}_{XY}) \ge \nonumber \\
	&\hspace{8em}  e^{-n\left(\theta - \zeta_{\theta}(n) - g(\hat{P}_{\xv,\yv}) + g(\tilde{P}_{XY}) \right)} \bigg)\d \theta.  \label{eq:perror-xy-lower-bound}
\end{align}
Using similar steps, one also finds
\begin{align}
	&\Pr\left( \error \mgiven \xv,\yv \right) \nonumber\\
	&\hspace{0.5em}\le nL e^{n\delta(n)} \max_{\tilde{P}_{XY}} \int_{0}^{\infty} e^{-nL\theta} \cdot \Pbb\bigg( N_n(\tilde{P}_{XY}) \ge \nonumber\\
	&\hspace{8.5em} e^{-n\left(\theta - g(\hat{P}_{\xv,\yv}) + g(\tilde{P}_{XY}) + \delta(n) \right)} \bigg)\d \theta. \label{eq:perror-xy-upper-bound}
\end{align}

The task then becomes to study the integrals that appear in~\eqref{eq:perror-xy-lower-bound} and \eqref{eq:perror-xy-upper-bound}. It can be shown that the asymptotic exponent of both of them is
\begin{align*}
	&\hat{E}_a(\hat{P}_{\xv,\yv},\tilde{P}_{XY})
	\coloneqq \left[ I_{\tilde{P}}(X \colon Y) - R \right]_+ \nonumber\\
	&\hspace{3em} + L \left[ g(\hat{P}_{\xv,\yv}) - g(\tilde{P}_{XY}) - \left[ R - I_{\tilde{P}}(X \colon Y) \right]_+ \right]_+.
\end{align*}
A detailed proof of this result is presented in the Appendix; here we overview key ideas. The random variable $N_n(\tilde{P}_{XY})$ follows a binomial distribution with $M-1 \doteq e^{nR}$ trials and probability of success $\Pbb \big( (\bar{\Xv},\yv) \in \Tcal_n(\tilde{P}_{XY}) \big) \doteq e^{-I_{\tilde{P}}(X \colon Y)}$, and its right tail has exponent $\Pbb\big( N_n(\tilde{P}_{XY}) \ge e^{nC} \big) \doteq e^{-nE_t}$, for $C\in\R$, where \cite[Thm.~4.1]{merhav2025}
\begin{align*}
	E_t
	\coloneqq
	\begin{cases}
		\left[I_{\tilde{P}}(X \colon Y)-R\right]_+, & \left[R-I_{\tilde{P}}(X \colon Y)\right]_+ \ge C,\\
		\infty, & \text{otherwise}.
	\end{cases}
\end{align*}
We can show that the exponent of the integrals are not affected by the vanishing terms $\zeta_\theta(n)$ and $\delta(n)$, and that they are the same as if we took the exponent of the right tail probability with $C = g(\hat{P}_{\xv,\yv}) - g(\tilde{P}_{XY}) - \theta$ and replaced it in the integral (see also \cite[p.~5042]{merhav2017}). See the Appendix for a full derivation\footnote{
	In proving the exponents of~\eqref{eq:perror-xy-lower-bound} and \eqref{eq:perror-xy-upper-bound}, we needed to use the fact that $L$ is fixed, so this proof technique is not immediately applicable to the exponential list size. A different technique is adopted in Section~\ref{sec:exponential-list-size}.
}.

With this result, we have
\begin{equation*}
	\Pr\left( \error \mgiven \xv,\yv \right)
	\doteq e^{-n \min_{\tilde{P}_{XY}} \hat{E}_a(\hat{P}_{\xv,\yv},\tilde{P}_{XY})}.
\end{equation*}
Replacing in~\eqref{eq:prob-error-L-0}, and applying standard steps using the method of types, we get
\begin{align*}
	\bar{p}_{e}(n,L)
	&\doteq \sum_{P_{XY} \colon P_X=Q_n} e^{-nD\left( P_{XY} \| Q \times W \right)}\\
	&\hspace{7em}\times e^{-n \min_{\tilde{P}_{XY}} \hat{E}_a({P}_{XY},\tilde{P}_{XY})}\\
	&\doteq e^{-n \min_{P_{XY}} \min_{\tilde{P}_{XY}} \left( D\left( P_{XY} \| Q \times W \right) + \hat{E}_a({P}_{XY},\tilde{P}_{XY}) \right) }.
\end{align*}
This concludes the sketch of the proof.

\section{Exponential List Size} \label{sec:exponential-list-size}

\subsection{Main Result and Discussion}

Our main result in the exponential list size regime $L(n) = e^{n\lambda}$, with\footnote{
	Differently from deterministic list decoding, here, the value~$\lambda$ of the list size is not limited above. Indeed, due to the randomised nature of the list, not even $\lambda \ge R$ can ensure that the correct message will be included in the list.
} $\lambda>0$, is a lower bound for the error exponent.

\begin{theorem}[Mismatched metric] \label{thm:error-exponent-exponential-L-mismatched}
	Under constant-composition random coding with distribution~$Q$, the average error probability of randomised list decoding with decoding metric $u_n(\xv,\yv) = e^{ng(\hat{P}_{\xv,\yv})}$ and list size $L(n) = e^{n\lambda}$, $\lambda\ge0$, satisfies $\lim_{n\to\infty} -\frac{1}{n}\log \bar{p}_{e}\big(n,\lfloor e^{n\lambda} \rfloor\big) \ge E_2(R,Q,\lambda)$,
	where
	\begin{align} \label{eq:error-exponent-exponential-L-mismatched}
		&E_2(R,Q,\lambda) \coloneqq \nonumber\\
		&\hspace{0.3em}\min_{P_{XY} \colon P_X=Q} \min_{\tilde{P}_{XY} \colon \tilde{P}_X=Q, \tilde{P}_Y = P_Y} D(P_{XY} \| Q \times W) \nonumber\\
		&\hspace{0.3em}+ \left[ I_{\tilde{P}}(X \colon Y) - (R - \lambda) -\big( g(\tilde{P}_{XY}) - g(P_{XY}) \big) \right]_+.
	\end{align}
\end{theorem}

\begin{corollary}[Matched metric] \label{cor:error-exponent-exponential-L-matched}
	For the matched metric $g(P_{XY}) = \E_{P}\left[ \log W(Y \given X) \right]$, the error exponent lower bound~\eqref{eq:error-exponent-exponential-L-mismatched} becomes
	\begin{align} \label{eq:error-exponent-exponential-L-matched}
		&E_2(R,Q,\lambda)
		= E_{\r}(R-\lambda,Q) = \nonumber\\
		& \min_{P_{XY} \colon P_X=Q}
		D(P_{XY} \| Q \times W) 
		+ \left[ I_{P}(X \colon Y) - (R - \lambda) \right]_+.
	\end{align}
	Moreover, the same result is obtained with the MMI metric $g(P_{XY}) = I_{P}(X \colon Y)$
\end{corollary}
\begin{IEEEproof}
	In the matched case, we have $g(P_{XY}) = \E_{P}\left[ \log W(Y \given X) \right] = I_P(X \colon Y) - H_P(Y)$; replacing in~\eqref{eq:error-exponent-exponential-L-mismatched}, and using that $H_P(Y) =  H_{\tilde{P}}(Y)$ results in~\eqref{eq:error-exponent-exponential-L-matched}. In the MMI case, the result follows by direct substitution in~\eqref{eq:error-exponent-exponential-L-mismatched}.
\end{IEEEproof}

In the case of fixed list size $L$, Corollary~\ref{cor:error-exponent-exponential-L-matched} recovers Corollary~\ref{cor:error-exponent-fixed-L-matched} by setting $\lambda=0$. And, whenever $\lambda>0$, the lower bound~\eqref{eq:error-exponent-exponential-L-matched} is a strict improvement with respect to the trivial lower bound $E_{\r}(R,Q)$ given by ordinary decoding.

This result should be compared to its deterministic counterpart. The error exponent of deterministic list decoding in the exponential list size regime under matched metric is~\cite{merhav2017}
\begin{align}
	\tilde{E}_2(R,Q,\lambda) = E_{\sp}(R - \lambda),
\end{align}
and is optimal~\cite{shannon1967}.
In general, $\tilde{E}_2(R,Q,\lambda) \ge E_{2}(R,Q,\lambda)$, and they coincide for $R \ge R_0 + \lambda$, cf.~Section~\ref{subsec:classical-error-exponents}. Thus, in this regime of high rates, the exponent of randomised list decoding with exponential list size is optimal as well. Another consequence of Corollary~\ref{cor:error-exponent-exponential-L-matched} is that rates up to $R + \lambda$ are achievable (in the sense of list decoding) with randomised list decoding, in the exponential list size regime. Error exponents (and bounds) for exponential list size are illustrated in Fig.~\ref{fig:example} on a BSC (green curves).

\subsection{Proof of Theorem~\ref{thm:error-exponent-exponential-L-mismatched}} \label{subsec:proof-exponential-L}

\begin{lemma}
	Let $X$ be a random variable with support in $\left[0,1\right]$ and $L>0$, then
	\begin{align} \label{eq:auxiliary-inequality-x-to-L}
		\E\left[ X^L \right]
		\le \min_{\xi \in \left[0,1\right]} \left\{ \xi^{L-1} \E\left[ X \right] + \Pbb\left( X > \xi \right) \right\}.
	\end{align}
\end{lemma}
\begin{IEEEproof}
	Fix $\xi \in \left[0,1\right]$ and split $\E\left[ X^L \right] = \E\left[ X^L \1\left\{ X \le \xi \right\} \right] + \E\left[ X^L \1\left\{ X > \xi \right\} \right]$.
	Bound the first term as $\E\left[ X^L \1\left\{ X \le \xi \right\} \right] \le \xi^{L-1} \E\left[ X \1\left\{ X<\xi \right\} \right] \le \xi^{L-1} \E\left[ X \right]$.
	For the second term, since $X \in \left[0,1\right]$, we have $\E\left[ X^L \1\left\{ X > \xi \right\} \right] \le \E\left[ \1\left\{ X > \xi \right\} \right] = \Pbb \left( X > \xi \right)$.
	Combine the two bounds and minimise over $\xi \in \left[ 0,1\right]$ to conclude.
\end{IEEEproof}

Note that we can apply \eqref{eq:auxiliary-inequality-x-to-L} to~\eqref{eq:list-decoding-error-probability}. Doing so, we have
\addtocounter{equation}{1}
\begin{align}
	\bar{p}_{e}(n,L)
	\le
	\min_{0 \le \xi \le 1}
	&\Bigg\{
	\underbrace{\xi^{L-1} \E\left[ 1 - \frac{u_n(\Xv_1, \Yv)}{\sum_{m'=1}^{M} u_n(\Xv_{m'}, \Yv)} \right]}_{\text{(\theequation a)}} \nonumber\\
	&+ \underbrace{\Pbb\left( 1 - \frac{u_n(\Xv_1, \Yv)}{\sum_{m'=1}^{M} u_n(\Xv_{m'}, \Yv)} > \xi \right)}_{\text{(\theequation b)}} \nonumber
	\Bigg\}.
\end{align}
The expectation in the first term is simply $\bar{p}_{e}(n,1)$, which can first be bounded using~\cite[Thm.~1]{scarlett2015} with $s=1$:
\begin{align*}
	\text{(\theequation a)}
	&\le \xi^{L-1} \E\left[
	\min\left\{1,\ 
	\frac{(M-1) \E\left[ u_n(\overline{\Xv},\Yv) \mgiven \Yv \right]}{u_n(\Xv_1,\Yv)}
	\right\}
	\right]\\
	&\le \E\left[
	\min\left\{1,\ 
	\xi^{L-1} \frac{(M-1) \E\left[ u_n(\overline{\Xv},\Yv) \mgiven \Yv \right]}{u_n(\Xv_1,\Yv)}
	\right\}
	\right],
\end{align*}
where $(\Xv_1,\bar{\Xv},\Yv) \sim P_{\Xv}(\xv) P_{\Xv}(\bar{\xv}) W^n(\yv \given \xv)$.
The second term can be rewritten and bounded as follows:
\begin{align*}
	\text{(\theequation b)}
	&= \E\left[ \Pbb\left( 1 - \frac{u_n(\Xv_1, \Yv)}{\sum_{m'=1}^{M} u_n(\Xv_{m'}, \Yv)} > \xi  \mgiven \Xv_1, \Yv \right) \right]\\ 
	&=\E \left[ \Pbb\left( \frac{{\sum_{m'\neq1}} u_n(\Xv_{m'}, \Yv)}{u_n(\Xv_1, \Yv)\frac{\xi}{1-\xi}}  > 1 \mgiven \Xv_1, \Yv \right) \right]\\ 
	&\le \E \left[ \min\left\{1, \ \frac{\E\left[{\textstyle\sum_{m'\neq1}} u_n(\Xv_{m'}, \Yv) \mgiven \Yv \right]}{u_n(\Xv_1, \Yv)\frac{\xi}{1-\xi}} \right\} \right]\\
	&= \E \left[ \min\left\{1, \ \frac{1-\xi}{\xi} \frac{(M-1) \E\left[u_n(\overline{\Xv}, \Yv) \mgiven \Yv \right]}{u_n(\Xv_1, \Yv)} \right\} \right],
\end{align*}
where the inequality comes from trivially bounding the probability by $1$, and applying Markov's inequality at the same time---whichever is better.

The bounds in (\theequation a) and (\theequation b) hold for any value $\xi \in \left[0,1\right]$; in particular, for the one satisfying $\xi^{L-1} = \frac{1-\xi}{\xi}$, which we denote $\xi^*(L)$, and that makes them coincide. With this choice of $\xi=\xi^{*}(L)$, we get
\begin{align} \label{eq:intermediate-bound-1}
	&\bar{p}_{e}(n,L)
	\le \nonumber \\
	&\quad 2 \, \E \left[\min\left\{1, \ \frac{1-\xi^*(L)}{\xi^*(L)} \frac{M \E\left[u_n(\overline{\Xv}, \Yv) \mgiven \Yv \right]}{u_n(\Xv_1, \Yv)} \right\} \right].
\end{align}

\begin{lemma}
	Let $L \ge 1$. The value $\xi^*(L)$ satisfying $\xi^{L-1} = \frac{1-\xi}{\xi}$ in the interval $\xi \in \left[0,1\right]$ is unique and satisfies
	\begin{equation} \label{eq:bounds-xi-star-L}
		1 - \frac{\log L}{L} \le \xi^*(L) \le \frac{L}{L+1},
	\end{equation}
	except that the lower bound is only claimed for $L \ge e$.
\end{lemma}
\begin{IEEEproof}
	The value $\xi^*(L)$ is the root of the function $f(\xi) = \xi^L + \xi - 1$. The function is strictly increasing, and has $f(0)<0$ and $f(1)>0$, and thus, a unique root.
	To derive the bounds, denote $x\coloneqq 1-\xi$, so that $\xi^L + \xi - 1 = 0 \iff x = (1-x)^L$.
	For the upper bound, $x = (1-x)^L \ge 1-Lx \iff x \ge \frac{1}{L+1}$, hence $\xi = 1-x \le \frac{L}{L+1}$.
	For the lower bound, note that $x = (1-x)^L \iff \log x = L \log(1-x)$. Using that $\log(1-x) \le -x$, we have $\log x \le -Lx \iff xe^{Lx} \le 1 \iff x\le W(L)/L$, where $W$ is the Lambert $W$ function, which satisfies $W(L) \le \log L$, for $L\ge e$~\cite{hoorfar2008}. This yields $\xi = 1-x \ge 1-\log L/L$.
\end{IEEEproof}

Noting that $\xi \mapsto \frac{1-\xi}{\xi}$ in~\eqref{eq:intermediate-bound-1} is decreasing, and applying the lower bound of~\eqref{eq:bounds-xi-star-L} with $L(n) = e^{n\lambda}$, we get
\begin{align}
	\frac{1-\xi^*(L)}{\xi^*(L)}
	\le \frac{\log L(n)}{L(n) - \log L(n)}
	= \frac{n\lambda}{e^{n\lambda} - n\lambda}
	\doteq e^{-n\lambda}. \label{eq:bound-L-factor}
\end{align}

We use the method of types to compute the inner expectation in~\eqref{eq:intermediate-bound-1}, for fixed $\Yv=\yv$:
\begin{align}
	\E\left[ u_n(\bar{\Xv},\yv) \right]
	&= \sum_{\bar{\xv}\in\Xcal^n} P_{\Xv}(\bar{\xv}) e^{ng(\hat{P}_{\bar{\xv},\yv})} \nonumber\\
	&= \sum_{\tilde{P}_{XY} \colon \tilde{P}_X=Q_n, \tilde{P}_Y=\hat{P}_{\yv}} \Pbb\left( (\bar{\Xv},\yv) \in \Tcal_n(\tilde{P}_{XY}) \right) \nonumber\\
	&\hspace{9em} \times  e^{ng(\tilde{P}_{XY})} \nonumber\\
	&\doteq \max_{\tilde{P}_{XY} \colon \tilde{P}_X=Q_n, \tilde{P}_Y=\hat{P}_{\yv}} e^{-nI_{\tilde{P}}(X \colon Y)} e^{ng(\tilde{P}_{XY})} \nonumber\\
	&= e^{-n \min_{\tilde{P}_{XY} \colon \tilde{P}_X=Q_n, \tilde{P}_Y=\hat{P}_{\yv}} \left( I_{\tilde{P}}(X \colon Y) - g(\tilde{P}_{XY}) \right)}. \label{eq:bound-inner-expectation}
\end{align}
For notational convenience, define
\begin{equation*}
	\hat{E}_{b}(Q,P_Y) \coloneqq \min_{\tilde{P}_{XY} \colon \tilde{P}_X=Q, \tilde{P}_Y=P_Y} I_{\tilde{P}}(X \colon Y) - g(\tilde{P}_{XY}).
\end{equation*}

Replacing \eqref{eq:bound-L-factor} and \eqref{eq:bound-inner-expectation} in \eqref{eq:intermediate-bound-1}, and using standard steps from the method of types yields
\begin{align*}
	\bar{p}_{e}(n,L(n))
	&\dotle \sum_{\xv \in \Tcal_n(Q_n)} \sum_{\yv\in\Ycal^n}
	P_{\Xv}(\xv) W^n(\yv \given \xv)\\
	&\hspace{2em} \times \min\left\{e^{n0}, \ e^{n\lambda} \frac{e^{nR} \cdot e^{-n\hat{E}_b(Q_n,\hat{P}_{\yv})}}{e^{ng(\hat{P}_{\xv,\yv})}} \right\}\\
	&\doteq \sum_{P_{XY} \colon P_X=Q_n} e^{-nD(P_{XY} \| Q \times W)}\\
	&\hspace{2em}\times \min\left\{ e^{n0},\ e^{ - n \left( \hat{E}_b(Q_n,P_Y) - (R - \lambda) + g(P_{XY}) \right) }  \right\}\\
	&\doteq \max_{P_{XY} \colon P_X=Q_n} e^{-nD(P_{XY} \| Q \times W)}\\
	&\hspace{2em} \times e^{ - n \left[ \hat{E}_b(Q_n,P_Y) - (R - \lambda) + g(P_{XY}) \right]_+ }\\
	&\doteq e^{ -n E_2(R,Q,\lambda)}.
\end{align*}
This concludes the proof.

\cleardoublepage
\bibliographystyle{IEEEtran}
\bibliography{references}

\clearpage
\newpage
\appendix

In this appendix we provide a detailed proof of Theorem~\ref{thm:error-exponent-fixed-L-mismatched}. The proof is an extension of the technique of~\cite{merhav2017} (see also~\cite[Sec.~4.4.3]{merhav2025}) to fixed list size $L \ge 1$. The tools we use are essentially the same as used in~\cite[Appendix~D]{merhav2025}.

The probability of error~\eqref{eq:list-decoding-error-probability} can be written as
\begin{align} \label{eq:prob-error-fixed-L}
	\bar{p}_e(n,L)
	= \sum_{\xv\in\Xcal^n} \sum_{\yv\in\Ycal^n} P_{\Xv}(\xv) W^n(\yv \given \xv) \Pr\left( \error \given \xv,\yv \right),
\end{align}
with
\begin{align*}
	&\Pr\left( \error \given \xv,\yv \right)\\
	&\hspace{1em}= \E\left[ \left( \frac{\sum_{m'\neq1} u_n(\Xv_{m'},\yv)}{u_n(\xv,\yv) + \sum_{m'\neq1} u_n(\Xv_{m'},\yv)} \right)^L \right]\\
	&\hspace{1em}= \E\left[ \left( \frac{\sum_{\tilde{P}_{XY}} N_n(\tilde{P}_{XY}) e^{ng(\tilde{P}_{XY})}}{e^{ng(\hat{P}_{\xv,\yv})} + \sum_{\tilde{P}_{XY}} N_n(\tilde{P}_{XY}) e^{ng(\tilde{P}_{XY})}} \right)^L \right],
\end{align*}
where the random variable
\begin{align} \label{eq:def-type-class-enumerator}
	N_n(\tilde{P}_{XY})
	&\coloneqq N_n(\tilde{P}_{XY};\Xv_2,\dots,\Xv_M,\yv) \nonumber\\
	&\coloneqq \sum_{m'=2}^{M} \1_{\Tcal_n(\tilde{P}_{XY})}(\Xv_{m'},\yv),
\end{align}
is a \emph{type class enumerator}~\cite[Ch.~4]{merhav2025}; and the sums are over the set
\begin{align*}
	\Vcal_n(\hat{P}_{\xv,\yv}) \coloneqq \left\{ \tilde{P}_{XY} \in \Pcal_n(\Xcal \times \Ycal) \colon \tilde{P}_X = \hat{P}_{\xv},\ \tilde{P}_Y = \hat{P}_{\yv} \right\}.
\end{align*}

Using the integral representation $\E[X] = \int_0^1 \Pbb\left( X\ge x \right)\d x$ for the expectation of a random variable $X \in \left[0,1\right]$, the change of variables $e^{-nL\theta} = t$, and simple algebraic manipulation, we find
\begin{align*}
	&\Pr\left( \error \given \xv,\yv \right) \nonumber\\
	&\hspace{0.2em}= \int_0^1 \Pbb\left( \frac{
		\sum_{\tilde{P}_{XY}} N_n(\tilde{P}_{XY}) e^{ng(\tilde{P}_{XY})}
	}{
		e^{ng(\hat{P}_{\xv,\yv})}
		+ \sum_{\tilde{P}_{XY}} N_n(\tilde{P}_{XY}) e^{ng(\tilde{P}_{XY})}
	} \ge t^{\frac{1}{L}} \right) \d t \nonumber\\
	&\hspace{0.2em}= nL \int_0^\infty e^{-nL\theta} \nonumber\\
	&\hspace{1.2em}\times\Pbb \left( \frac{
			\sum_{\tilde{P}_{XY}} N(\tilde{P}_{XY}) e^{ng(\tilde{P}_{XY})}
		}{
			e^{ng(\hat{P}_{\xv,\yv})}
			+ \sum_{\tilde{P}_{XY}} N(\tilde{P}_{XY}) e^{ng(\tilde{P}_{XY})}
		} \ge e^{-n\theta} \right) \d \theta \nonumber\\
	&\hspace{0.2em}= nL \int_0^\infty e^{-nL\theta}\\
	&\hspace{1.2em} \times\Pbb \left(
	\sum_{\tilde{P}_{XY}} N(\tilde{P}_{XY}) e^{ng(\tilde{P}_{XY})}
	\ge \frac{e^{-n\theta}}{1-e^{-n\theta}} e^{ng(\hat{P}_{\xv,\yv})} \right) \d \theta \nonumber.
\end{align*}

In the following, denote $\delta(n) \coloneqq \frac{|\Xcal||\Ycal|}{n} \log(n+1)$, and recall that the number of types $\tilde{P}_{XY}$ over which the sum acts is upper bounded by $\big| \Vcal_n(\hat{P}_{\xv,\yv}) \big| \le e^{n\delta(n)}$. Moreover, denote $\zeta_\theta(n) \coloneqq -\frac{1}{n}\log \left( 1-e^{-n\theta} \right)$.
First, we compute the lower bound
\newpage
\begin{align}
	&\Pr\left( \error \mgiven \xv,\yv \right) \nonumber\\
	&= nL \int_{0}^{\infty} e^{-nL\theta} \nonumber\\
	&\hspace{2.5em}\times \Pbb\left( \sum_{\tilde{P}_{XY}} N_n(\tilde{P}_{XY}) e^{ng(\tilde{P}_{XY})} \ge e^{-n\left(\theta - \zeta_{\theta}(n) - g(\hat{P}_{\xv,\yv}) \right)} \right)\d \theta \nonumber\\
	&\ge nL \int_{0}^{\infty} e^{-nL\theta} \nonumber\\
	&\hspace{2.5em} \times \max_{\tilde{P}_{XY}} \Pbb\left( N_n(\tilde{P}_{XY}) e^{ng(\tilde{P}_{XY})} \ge e^{-n\left(\theta - \zeta_{\theta}(n) - g(\hat{P}_{\xv,\yv}) \right)} \right)\d \theta \nonumber\\
	&\ge nL \int_{0}^{\infty} e^{-nL\theta} e^{-n\delta(n)} \nonumber\\
	&\hspace{2.5em}\times \sum_{\tilde{P}_{XY}} \Pbb\left( N_n(\tilde{P}_{XY}) e^{ng(\tilde{P}_{XY})} \ge e^{-n\left(\theta - \zeta_{\theta}(n) - g(\hat{P}_{\xv,\yv}) \right)} \right)\d \theta \nonumber\\
	&= nL e^{-n\delta(n)} \sum_{\tilde{P}_{XY}} \int_{0}^{\infty} e^{-nL\theta} \nonumber\\
	&\hspace{3.5em} \times \Pbb\left( N_n(\tilde{P}_{XY}) e^{ng(\tilde{P}_{XY})} \ge e^{-n\left(\theta - \zeta_{\theta}(n) - g(\hat{P}_{\xv,\yv}) \right)} \right)\d \theta \nonumber\\
	&\ge nL e^{-n\delta(n)} \max_{\tilde{P}_{XY}} \int_{0}^{\infty} e^{-nL\theta} \nonumber\\
	&\hspace{3.5em} \times \Pbb\left( N_n(\tilde{P}_{XY}) e^{ng(\tilde{P}_{XY})} \ge e^{-n\left(\theta - \zeta_{\theta}(n) - g(\hat{P}_{\xv,\yv}) \right)} \right)\d \theta \nonumber\\
	&= nL e^{-n\delta(n)} \max_{\tilde{P}_{XY}} \int_{0}^{\infty} e^{-nL\theta} \nonumber \\
	&\hspace{3.5em} \times \Pbb\left( N_n(\tilde{P}_{XY}) \ge e^{-n\left(\theta - \zeta_{\theta}(n) - g(\hat{P}_{\xv,\yv}) + g(\tilde{P}_{XY}) \right)} \right)\d \theta\nonumber.
\end{align}
We also have the upper bound
\begin{align}
	&\Pr\left( \error \mgiven \xv,\yv \right) \nonumber\\
	&= nL \int_{0}^{\infty} e^{-nL\theta} \nonumber\\
	&\hspace{0.5em} \times \Pbb\left( \sum_{\tilde{P}_{XY}} N_n(\tilde{P}_{XY}) e^{ng(\tilde{P}_{XY})} \ge e^{-n\left(\theta - \zeta_{\theta}(n) - g(\hat{P}_{\xv,\yv}) \right)} \right)\d \theta \nonumber\\
	&\le nL \int_{0}^{\infty} e^{-nL\theta} \cdot
	\Pbb\Bigg( e^{n\delta(n)} \max_{\tilde{P}_{XY}} N_n(\tilde{P}_{XY}) e^{ng(\tilde{P}_{XY})} \ge \nonumber\\
	&\hspace{13em} e^{-n\left(\theta - \zeta_{\theta}(n) - g(\hat{P}_{\xv,\yv}) \right)} \Bigg) \d \theta \nonumber\\
	&\le nL \int_{0}^{\infty} e^{-nL\theta} \sum_{\tilde{P}_{XY}} \Pbb\bigg( N_n(\tilde{P}_{XY}) e^{ng(\tilde{P}_{XY})} \ge \nonumber\\
	&\hspace{11em}  e^{-n\left(\theta - \zeta_{\theta}(n) - g(\hat{P}_{\xv,\yv}) + \delta(n) \right)} \bigg)\d \theta \nonumber\\
	&= nL \sum_{\tilde{P}_{XY}} \int_{0}^{\infty} e^{-nL\theta} \cdot \Pbb\bigg( N_n(\tilde{P}_{XY}) e^{ng(\tilde{P}_{XY})} \ge \nonumber\\
	&\hspace{11em}  e^{-n\left(\theta - \zeta_{\theta}(n) - g(\hat{P}_{\xv,\yv}) + \delta(n) \right)} \bigg)\d \theta \nonumber\\
	&\le nL e^{n\delta(n)} \max_{\tilde{P}_{XY}} \int_{0}^{\infty} e^{-nL\theta} \cdot \Pbb\bigg( N_n(\tilde{P}_{XY}) e^{ng(\tilde{P}_{XY})} \ge \nonumber\\
	&\hspace{11em}  e^{-n\left(\theta - \zeta_{\theta}(n) - g(\hat{P}_{\xv,\yv}) + \delta(n) \right)} \bigg)\d \theta \nonumber\\
	&\le nL e^{n\delta(n)} \max_{\tilde{P}_{XY}} \int_{0}^{\infty} e^{-nL\theta} \nonumber\\
	&\hspace{0.5em} \times \Pbb\left( N_n(\tilde{P}_{XY}) \ge  \cdot e^{-n\left(\theta - g(\hat{P}_{\xv,\yv}) + g(\tilde{P}_{XY}) + \delta(n) \right)} \right)\d \theta\nonumber,
\end{align}
where in the last step we used that $\zeta_{\theta}(n) > 0$.
Together, we get, for each $n\in\N$,
\begin{align} \label{eq:lower-upper-bound-p-error-xy}
	&nL e^{-n\delta(n)} \max_{\tilde{P}_{XY} \in \Vcal_n(\hat{P}_{\xv,\yv})} \int_{0}^{\infty} e^{-nL\theta} \nonumber\\
	&\hspace{1em}\times \Pbb\left( N_n(\tilde{P}_{XY}) \ge e^{n \left( g(\hat{P}_{\xv,\yv}) - g(\tilde{P}_{XY}) - \theta + \zeta_\theta(n) \right)} \right)\d \theta \nonumber\\
	&\le
	\Pr\left( \error \mgiven \xv,\yv \right) \nonumber\\
	&\le nL e^{n\delta(n)} \max_{\tilde{P}_{XY} \in \Vcal_n(\hat{P}_{\xv,\yv})} \int_{0}^{\infty} e^{-nL\theta} \nonumber\\
	&\hspace{1em} \times \Pbb\left( N_n(\tilde{P}_{XY}) \ge e^{n \left( g(\hat{P}_{\xv,\yv}) - g(\tilde{P}_{XY}) - \theta - \delta(n) \right)} \right)\d \theta.
\end{align}

The task is now to study the integrals\footnote{
	The results are a reminiscent of Laplace method~(e.g., \cite[Ch.~3]{merhav2025}) or Varadhan's theorem~(e.g., \cite[Sec.~4.3]{dembo1998}), except that the exponent of the function inside the integral depends on $n$, cf.~\cite[Thm.~4.1]{merhav2025}. This, together with the fact that the convergence of that function is not uniform with respect to the integrand~$\theta$, prevents direct application of those results, at least as they are stated.
}. We will do that separately for upper and lower bounds. We will use that the \emph{type class enumerator}~\eqref{eq:def-type-class-enumerator} follows a binomial distribution~\cite[Ch.~4]{merhav2025}
\begin{align*}
	N_n(P_{XY}) \sim \Bin\left( M-1,\ \Pbb\left( (\bar{\Xv},\yv) \in \Tcal_n(P_{XY}) \right) \right),
\end{align*}
with
\begin{equation*}
	M-1 = e^{n\left( R - \epsilon_1(n) \right)}
\end{equation*}
trials, and probability of success
\begin{equation*}
	\Pbb\left( (\Xv,\yv) \in \Tcal_n(\tilde{P}_{XY}) \right) = e^{-n\left( I_{\tilde{P}}(X \colon Y) - \epsilon_2(n) \right)},
\end{equation*}
where
\begin{equation*}
	\epsilon_1(n) = \frac{1}{n} \log \frac{e^{nR}}{e^{nR}-1} > 0
\end{equation*}
and
\begin{equation*}
	\left| \epsilon_2(n) \right| \le \frac{|\Xcal||\Ycal|}{n} {\log(n+1)}.
\end{equation*}

\subsection{Lower Bound}
Let us start with the integral in the lower bound, namely,
\begin{align} \label{eq:lower-bound-integral}
	I_{\mathrm{lb}}(n)
	&\coloneqq \int_{0}^{\infty} e^{-nL\theta} \cdot \Pbb\bigg( N_n(\tilde{P}_{XY}) \ge \nonumber\\
	&\hspace{6em} e^{n \left( g(\hat{P}_{\xv,\yv}) - g(\tilde{P}_{XY}) - \theta + \zeta_\theta(n) \right)} \bigg)\d \theta.
\end{align}
For convenience, we shall denote
\begin{align*}
	\lambda(n,\theta) \coloneqq g(\hat{P}_{\xv,\yv}) - g(\tilde{P}_{XY}) - \theta + \zeta_\theta(n).
\end{align*}
Before proceeding, we present an auxiliary result.

\begin{proposition} \label{prop:function-fn-zeta}
	For each $n\in\N$, consider the function $f_n \colon \theta \mapsto \theta - \zeta_{\theta}(n)$, with $\zeta_\theta(n) = -\frac{1}{n}\log \left( 1-e^{-n\theta} \right)$. For any $a\in\R$, there exists a value $\theta_n^{\star} \coloneqq \theta_n^{\star}(a)$ such that $\theta > \theta_n^{\star} \implies f_n(\theta) > a$. This value is given by
	\begin{align}
		\theta_n^{\star}(a) = \frac{1}{n} \log\left( e^{na} + 1 \right).
	\end{align}
	And, if $a\ge0$, then $0< \theta_n^{\star}(a) - a \le \frac{\log 2}{n}$.
\end{proposition}
\begin{IEEEproof}
	The function $f_n$ is increasing, for its derivative is $f_n'(\theta) = e^{n\theta}/(e^{n\theta}-1) > 0$. Its extreme values are $\lim_{\theta\to0^+} f_n(\theta) = -\infty$, and $\lim_{\theta\to\infty} f_n(\theta) = \infty$, so that $\Im f = \R$. The value $\theta_n^{\star}(a)$ is the (only) root of the function $F_n(\theta) \coloneqq \theta + \frac{1}{n} \log \left( 1 - e^{-n\theta} \right) - a$, which can be computed in closed form, and solves to $\theta_n^{\star}(a) = \frac{1}{n} \log\left( e^{na} + 1 \right)$.
	Due to the term $\zeta_{\theta}(n) > 0$ that intervenes in $f_n$, $\theta_n^{\star}$ is larger than~$a$. The difference is precisely $\theta_n^{\star}\left(a\right) - a = \frac{1}{n} \log \left( \frac{e^{na} + 1}{e^{na}} \right)$; if $a \ge 0$, it satisfies $0 < \frac{1}{n} \log \left( \frac{e^{na} + 1}{e^{na}} \right) \le \frac{\log 2}{n}$.
\end{IEEEproof}

Next, we will split the proof in two cases: $R < I_{\tilde{P}}(X \colon Y)$ and $R \ge I_{\tilde{P}}(X \colon Y)$.

\textit{Case $R < I_{\tilde{P}}(X \colon Y)$:} note that, if $\lambda(n,\theta) < 0 \iff e^{n\lambda(n,\theta)} \le 1$, then
\begin{align}
	&\hspace{-1em}\Pbb\left( N_n(\tilde{P}_{XY}) \ge e^{n\lambda(n,\theta)} \right) \nonumber\\
	&\hspace{1em}\ge \Pbb\left( N_n = 1 \right)
	\nonumber\\
	&\hspace{1em}= \binom{e^{n\left( R - \epsilon_1(n) \right)}}{1}
	e^{-n\left( I_{\tilde{P}}(X \colon Y) - \epsilon_2(n) \right)} \nonumber\\
	&\hspace{2em}\times \left( 1 - e^{-n\left( I_{\tilde{P}}(X \colon Y) - \epsilon_2(n) \right)} \right)^{e^{n\left( R - \epsilon_1(n) \right)}-1}
	\nonumber\\
	&\hspace{1em}= e^{-n\left( I_{\tilde{P}}(X \colon Y) - R + \epsilon_1(n) - \epsilon_2(n) - \Delta_{\tilde{P}}(n) \right)},
	\label{eq:lower-bound-case1}
\end{align}
where
\begin{equation*}
	\Delta_{\tilde{P}}(n)
	\coloneqq \frac{1}{n} \log \left[ \left( 1 - e^{-n\left( I_{\tilde{P}}(X \colon Y) - \epsilon_2(n) \right)} \right)^{e^{n\left( R - \epsilon_1(n) \right)}-1} \right],
\end{equation*}
and $\Delta_{\tilde{P}}(n) \to 0$, for each $\tilde{P}_{XY}$ such that $R < I_{\tilde{P}}(X \colon Y)$---but not uniformly.
The condition for~\eqref{eq:lower-bound-case1} translates to
\begin{align*}
	\lambda(n,\theta) < 0
	&\iff \theta - \zeta_n(\theta) > g(\hat{P}_{\xv,\yv}) - g(\tilde{P}_{XY}).
\end{align*}
Thanks to Proposition~\ref{prop:function-fn-zeta}, there exists
\begin{align*}
	\theta_n^{\star} \coloneqq \frac{1}{n}\log \left( e^{n\left[ g(\hat{P}_{\xv,\yv})-g(\tilde{P}_{XY})  \right]_+} + 1 \right)
\end{align*}
such that
\begin{align*}
	\theta > \theta_n^{\star}
	\implies
	\theta - \zeta_n(\theta) > \left[ g(\hat{P}_{\xv,\yv}) - g(\tilde{P}_{XY}) \right]_+,
\end{align*}
so that the bound \eqref{eq:lower-bound-case1} holds in that interval. Moreover, denote $\beta^{\star}(n) \coloneqq \theta_n^{\star} - \big[ g(\hat{P}_{\xv,\yv}) - g(\tilde{P}_{XY}) \big]_+ \to 0$.
So we can bound the integral~\eqref{eq:lower-bound-integral} as follows:
\begin{align}
	&\hspace{-1em}I_{\mathrm{lb}}(n) \nonumber\\
	&\ge
	\int_{\theta_n^{\star}}^{\infty} e^{-nL\theta} \nonumber\\
	&\hspace{1em}\times \Pbb\left( N(\tilde{P}_{XY}) \ge e^{n\left(g(\hat{P}_{\xv,\yv}) - g(\tilde{P}_{XY}) -\theta + \zeta_\theta(n) \right) } \right)\d \theta \nonumber\\
	&\ge \int_{\theta_n^{\star}}^{\infty}
	e^{-nL\theta} \cdot
	e^{-n\left( I_{\tilde{P}}(X \colon Y) - R + \epsilon_1(n) - \epsilon_2(n) - \Delta_{\tilde{P}}(n) \right)}
	\d \theta \nonumber\\
	&= e^{-n\left( I_{\tilde{P}}(X \colon Y) - R + \epsilon_1(n) - \epsilon_2(n) - \Delta_{\tilde{P}}(n) \right)} 
	\frac{e^{-nL\theta_n^{\star}}}{nL} \nonumber\\
	&= \frac{1}{nL} e^{-n\left( I_{\tilde{P}}(X \colon Y) - R + \epsilon_1(n) - \epsilon_2(n) - \Delta_{\tilde{P}}(n) \right)} \nonumber\\
	&\hspace{1em}\times e^{-nL \left( \big[ g(\hat{P}_{\xv,\yv}) - g(\tilde{P}_{XY})\big]_+ + \beta^{\star}(n) \right) } \nonumber\\
	&\doteq e^{-n\left( I_{\tilde{P}}(X \colon Y) - R + L \big[ g(\hat{P}_{\xv,\yv}) - g(\tilde{P}_{XY})\big]_+ - \Delta_{\tilde{P}}(n) \right)}. \label{eq:result-lower-bound-1}
\end{align}

\textit{Case $R \ge I_{\tilde{P}}(X \colon Y)$:} observe that, if $e^{n\lambda(n,\theta)} \le \E\big[N_n(\tilde{P}_{XY})\big]$, then
\begin{equation} \label{eq:lower-bound-case2}
	\Pbb\left( N_n(\tilde{P}_{XY}) \ge e^{n\lambda(n,\theta)} \right) \ge \frac{1}{2}.
\end{equation}
This condition translates to
\begin{align*}
	&e^{n\lambda(n,\theta)} \le \E[N_n]= e^{n\left( R - I_{\tilde{P}}(X \colon Y) -\epsilon_1(n) + \epsilon_2(n) \right)}\\
	&\iff \theta - \zeta_\theta(n) \ge g(\hat{P}_{\xv,\yv}) - g(\tilde{P}_{XY}) - R + I_{\tilde{P}}(X \colon Y)\\
	&\hspace{8em} + \epsilon_1(n) - \epsilon_2(n).
\end{align*}
Again thanks to Proposition~\ref{prop:function-fn-zeta}, there exists
\begin{align*}
	&\theta_n^{\star\star} \coloneqq\\
	&\hspace{1em}\frac{1}{n}\log\left( e^{n\left[ g(\hat{P}_{\xv,\yv}) - g(\tilde{P}_{XY}) - R + I_{\tilde{P}}(X \colon Y) + \epsilon_1(n) - \epsilon_2(n) \right]_+} + 1 \right)
\end{align*}
such that
\begin{align*}
	\theta \ge \theta_n^{\star\star}
	\implies \theta - \zeta_\theta(n) \ge &\Big[ g(\hat{P}_{\xv,\yv}) - g(\tilde{P}_{XY}) - R\\
	&+ I_{\tilde{P}}(X \colon Y) + \epsilon_1(n) - \epsilon_2(n) \Big]_+,
\end{align*}
so that the bound \eqref{eq:lower-bound-case2} holds in that interval. Moreover, denote $\beta^{\star\star}(n) \coloneqq \theta_n^{\star\star} - \big[ g(\hat{P}_{\xv,\yv}) - g(\tilde{P}_{XY}) - R + I_{\tilde{P}}(X \colon Y) + \epsilon_1(n) - \epsilon_2(n) \big]_+ \to 0$.
With that, we can lower bound the integral~\eqref{eq:lower-bound-integral} in a similar way:
\begin{align}
	&\hspace{-1em}I_{\mathrm{lb}}(n) \nonumber\\
	&\ge
	\int_{\theta_n^{\star\star}}^{\infty} e^{-nL\theta} \nonumber\\
	&\hspace{1em} \times \Pbb\left( N(\tilde{P}_{XY}) \ge e^{n\left(g(\hat{P}_{\xv,\yv}) - g(\tilde{P}_{XY}) -\theta + \zeta_\theta(n) \right) } \right)\d \theta \nonumber\\
	&\ge
	\int_{\theta_n^{\star\star}}^{\infty} e^{-nL\theta} \left(\frac{1}{2}\right) \d \theta \nonumber\\
	&= \frac{1}{2} \frac{e^{-nL\theta_n^{\star\star}}}{nL} \nonumber\\
	&= \frac{1}{2nL} e^{-nL\left[ g(\hat{P}_{\xv,\yv}) - g(\tilde{P}_{XY}) - R + I_{\tilde{P}}(X \colon Y) + \epsilon_1(n) - \epsilon_2(n) \right]_+} \nonumber\\
	&\hspace{1em}\times e^{-nL\beta^{\star\star}(n)} \nonumber\\
	&\doteq e^{-nL\left( \left[ g(\hat{P}_{\xv,\yv}) - g(\tilde{P}_{XY}) - R + I_{\tilde{P}}(X \colon Y) \right]_+ \right)}. \label{eq:result-lower-bound-2}
\end{align}

\begin{remark}
	To get both \eqref{eq:result-lower-bound-1} and \eqref{eq:result-lower-bound-2}, we used the hypothesis that $L$ is fixed. Indeed, in the last step of each, we needed that $L \beta^{\star}(n) \to 0$, and $L \left( \epsilon_1(n) - \epsilon_2(n) + \beta^{\star\star}(n) \right) \to 0$, respectively. The largest of the vanishing terms is $\epsilon_2(n) = O\left( \frac{\log n}{n} \right)$.
	In fact, these derivations would still work if we let $L=L(n)$ grow with $n$, but limited to $L(n) = o\left( \frac{n}{\log n} \right)$. The factors $\frac{1}{nL}$ are not that important, as they cancel out with~\eqref{eq:lower-upper-bound-p-error-xy}.
\end{remark}

The results \eqref{eq:result-lower-bound-1} and \eqref{eq:result-lower-bound-2} can be unified as
\begin{align*}
	I_{\mathrm{lb}}(n) \dotge e^{-n\left( \hat{E}_{a}(\hat{P}_{\xv,\yv}, \tilde{P}_{XY}, R, L) - \Delta_{\tilde{P}}\1\left\{ R < I_{\tilde{P}}(X \colon Y) \right\} \right)},
\end{align*}
where, for convenience, we let
\begin{align} \label{eq:def-E3}
	&\hat{E}_{a}(P_{XY}, \tilde{P}_{XY}, R, L)
	\coloneqq
	\left[ I_{\tilde{P}}(X \colon Y) - R \right]_+\nonumber\\
	&\hspace{1em}+ L \left[ g(P_{XY}) - g(\tilde{P}_{XY}) - \left[ R - I_{\tilde{P}}(X \colon Y) \right]_+ \right]_+.
\end{align}

Replacing this result in~\eqref{eq:lower-upper-bound-p-error-xy}, we get
\begin{align}
	&\hspace{-0.7em}\Pr\left( \error \given \xv,\yv \right) \nonumber\\
	&\dotge \max_{\tilde{P}_{XY}}
	e^{-n\left( \hat{E}_{a}(\hat{P}_{\xv,\yv}, \tilde{P}_{XY}, R, L) - \Delta_{\tilde{P}}\1\left\{ R < I_{\tilde{P}}(X \colon Y) \right\} \right)} \nonumber\\
	&= 
	e^{-n \min_{\tilde{P}_{XY}} \left( \hat{E}_{a}(\hat{P}_{\xv,\yv}, \tilde{P}_{XY}, R, L) - \Delta_{\tilde{P}}\1\left\{ R < I_{\tilde{P}}(X \colon Y) \right\} \right)} \nonumber\\
	&\ge e^{-n \left( \min_{\tilde{P}_{XY}}
		\hat{E}_{a}(\hat{P}_{\xv,\yv}, \tilde{P}_{XY}, R, L)
		- \Gamma_{\hat{P}_{\xv,\yv}}(n) \right)}, \label{eq:lower-bound-for-tce}
\end{align}
where $\Gamma_{\hat{P}_{\xv,\yv}}(n) \coloneqq \Delta_{\tilde{P}^{\star}}(n)\1\left\{ R < I_{\tilde{P}^{\star}} \right\} \to 0$, with
\begin{align*}
	&\tilde{P}_{XY}^{\star} \coloneqq \min_{\tilde{P}_{XY} \in \Vcal_n(\hat{P}_{\xv,\yv})}  \left[ I_{\tilde{P}}(X \colon Y) - R \right]_+\\
	&\hspace{3em}+ L \left[ g(P_{XY}) - g(\tilde{P}_{XY}) - \left[ R - I_{\tilde{P}}(X \colon Y) \right]_+ \right]_+,
\end{align*}
and the dependency in $\hat{P}_{\xv,\yv}$ comes from the minimisation set $\Vcal_n(\hat{P}_{\xv,\yv})$.

Note that the asymptotic lower bound~\eqref{eq:lower-bound-for-tce} only depends on $\xv,\yv$ through their joint type~$\hat{P}_{\xv,\yv}$. Replacing that in~\eqref{eq:prob-error-fixed-L}, and applying standard steps from the method of types, we get the lower bound
\begin{align}
	&\bar{p}_{e}(n,L) \nonumber\\
	&\dotge \sum_{P_{XY} \colon P_X=Q_n}  e^{-nD(P_{XY}\| Q \times W)} \nonumber\\
	&\hspace{4em}\times e^{-n \left( \min_{\tilde{P}_{XY} \in \Vcal_n(P_{XY})}
		\hat{E}_{a}(P_{XY}, \tilde{P}_{XY}, R, L)
		- \Gamma_{P_{XY}}(n) \right)} \nonumber\\
	&\doteq \max_{P_{XY}\colon P_X=Q_n} e^{-nD(P_{XY}\| Q \times W)} \nonumber\\
	&\hspace{4em}\times e^{-n \left( \min_{\tilde{P}_{XY} \in \Vcal_n(P_{XY})} 
		\hat{E}_{a}(P_{XY}, \tilde{P}_{XY}, R, L)
		- \Gamma_{P_{XY}}(n) \right)} \nonumber\\
	&= e^{-n \min_{P_{XY}\colon P_X=Q_n} \min_{\tilde{P}_{XY} \in \Vcal_n(P_{XY})} \big(  D(P_{XY}\| Q \times W)} \nonumber\\
	&\hspace{12em}\phantom{e}^{ + \hat{E}_{a}(P_{XY}, \tilde{P}_{XY}, R, L)
		- \Gamma_{P_{XY}}(n) \big)} \nonumber\\
	&\ge e^{-n \Big( \min_{P_{XY}\colon P_X=Q_n} \min_{\tilde{P}_{XY} \in \Vcal_n(P_{XY})} \big(  D(P_{XY}\| Q \times W)} \nonumber\\
	&\hspace{12em}\phantom{e}^{ + \hat{E}_{a}(P_{XY}, \tilde{P}_{XY}, R, L) \big)
		- \Gamma_{P^{\star}_{XY}}(n)  \Big)} \nonumber
\end{align}
where
\begin{align*}
	P^{\star}_{XY}
	&\coloneqq
	\arg\min_{P_{XY}\in\Scal_n(Q)}
	\min_{\tilde{P}_{XY} \in \Vcal_n(P_{XY},Q)}
	D(P_{XY}\| Q \times W)\\
	&+ \hat{E}_{a}(P_{XY}, \tilde{P}_{XY}, R, L)
	- \Gamma_{P_{XY}}(n),
\end{align*}
and, in particular, $\Gamma_{P^{\star}_{XY}}(n) \to 0$. Recalling the definition~\eqref{eq:def-E3} and taking the limit, we conclude that
\begin{align*}
	&\lim_{n\to\infty}\frac{1}{n}\log \bar{p}_e(n,L)
	\le \min_{P_{XY} \colon P_X=Q}
	\min_{\tilde{P}_{XY} \colon \tilde{P}_X=Q,\ \tilde{P}_Y=P_Y}\\
	&\hspace{2em} D(P_{XY}\| Q \times W)
	+
	\left[ I_{\tilde{P}}(X \colon Y) - R \right]_+\\
	&\hspace{2em}+ L \left[ g(P_{XY}) - g(\tilde{P}_{XY}) - \left[ R - I_{\tilde{P}}(X \colon Y) \right]_+ \right]_+.
\end{align*}

\subsection{Upper Bound}

Now we want to deal with the integral in the upper bound, namely,
\begin{align}
	I_{\mathrm{ub}}(n)
	&\coloneqq \int_{0}^{\infty} e^{-nL\theta} \cdot \Pbb\Big( N_n(\tilde{P}_{XY}) \ge \nonumber\\
	&\hspace{6em} e^{n \left( g(\hat{P}_{\xv,\yv}) - g(\tilde{P}_{XY}) - \theta - \delta(n) \right)} \Big)\d \theta.
\end{align}
The result will follow by an application of Theorem~\ref{thm:integral}, whose proof is deferred to Section~\ref{subsec:appendix-proof-thm-integral} of the Appendix.

\begin{theorem} \label{thm:integral}
	Let $(A_n)_{n\in\N}$ and $(B_n)_{n\in\N}$ two positive sequences, with $A_n \to A$ and $B_n \to B$, and $(C_n)_{n\in\N}$ a real sequence with $C_n \to C$, uniformly in $\theta$, as $n\to\infty$. Let $\bL \in \N$ fixed, and $N_n \sim \Bin\left( e^{nA_n}, e^{-nB_n} \right)$. Then,
	\begin{align}
		&\int_0^{\infty} e^{-n\bL\theta} \cdot \Pbb\left( N_n \ge e^{n\left( C-\theta \right)} \right)\d\theta
		 \nonumber \\
		&\hspace{4em}\doteq e^{-n \left(
			\left[ B-A \right]_+  + \bL \left[ C - \left[ A - B \right]_+ \right]_+
			\right)}.
 	\end{align}
	In particular, when $\bL=1$, we have
	\begin{equation}
		\int_0^{\infty} e^{-n\theta} \cdot \Pbb\left( N_n \ge e^{n\left( C-\theta \right)} \right)\d\theta
		\doteq  e^{-n\left[ B - A + \left[ C \right]_+ \right]_+}.
	\end{equation}
\end{theorem}

Apply Theorem~\ref{thm:integral} with $A_n = R - \epsilon_1(n)$, $B_n = I_{\tilde{P}}(X \colon Y) - \epsilon_2(n)$, and $C_n = g(\hat{P}_{\xv,\yv}) - g(\tilde{P}_{XY}) - \theta - \delta(n)$.
Note that the convergence $C_n \to C$ is uniform in $\theta$, for $\delta(n) = \frac{|\Xcal||\Ycal|}{n} \log(n+1)$.
This yields
\begin{align*}
	&I_{\mathrm{ub}}(n) \doteq\\
	&\quad e^{-n \left( \left[ I_{\tilde{P}}(X \colon Y) - R \right]_+ + L \left[  I_{\tilde{P}}(X \colon Y) - R + \left[ g(\hat{P}_{\xv,\yv}) - g(\tilde{P}_{XY}) \right] \right]_+ \right)}.
\end{align*}

\begin{remark}
	Theorem~\ref{thm:integral} cannot be directly applied to lower bound~\eqref{eq:lower-bound-integral}, because of the term $\zeta_\theta(n)$ in that expression, that makes the convergence of that exponent non-uniform in~$\theta$.
\end{remark}

Replacing this result in~\eqref{eq:lower-upper-bound-p-error-xy}, and recalling the definition~\eqref{eq:def-E3}, we get
\begin{align*}
	\Pr\left( \error \given \xv,\yv \right)
	&\dotle \max_{\tilde{P}_{XY}\in\Vcal_n(\hat{P}_{\xv,\yv})}
	e^{-n \hat{E}_{a}(\hat{P}_{\xv,\yv},\tilde{P}_{XY},R,L)}\\
	&= 
	e^{-n \min_{\tilde{P}_{XY}\in\Vcal_n(\hat{P}_{\xv,\yv},Q)} \hat{E}_{a}(\hat{P}_{\xv,\yv},\tilde{P}_{XY},R,L)}.
\end{align*}
Again, noting that this asymptotic bound only depends on $\xv,\yv$ through the joint type $\hat{P}_{\xv,\yv}$, and replacing in~\eqref{eq:prob-error-fixed-L}, we get, with the method of types,
\begin{align*}
	\bar{p}_{e}(n,L)
	&\dotle \sum_{P_{XY} \colon P_X=Q_n} e^{-nD(P_{XY}\| Q \times W)}\\
	&\hspace{4em}\times e^{-n \min_{\tilde{P}_{XY} \in \Vcal_n(P_{XY},Q)}
		\hat{E}_{a}(P_{XY}, \tilde{P}_{XY}, R, L)
		} \\
	&\doteq \max_{P_{XY} \colon P_X=Q_n} e^{-nD(P_{XY}\| Q \times W)}\\
	&\hspace{4em}\times e^{-n \min_{\tilde{P}_{XY} \in \Vcal_n(P_{XY},Q)}
		\hat{E}_{a}(P_{XY}, \tilde{P}_{XY}, R, L)
		} \\
	&= e^{-n \min_{P_{XY}\colon P_X=Q_n} \min_{\tilde{P}_{XY} \in \Vcal_n(P_{XY})} \big(  D(P_{XY}\| Q \times W)} \nonumber\\
	&\hspace{12em}\phantom{e}^{ + \hat{E}_{a}(P_{XY}, \tilde{P}_{XY}, R, L) \big)  }.
\end{align*}
Thus, taking the limit,
\begin{align*}
	&\lim_{n\to\infty}\frac{1}{n}\log \bar{p}_e(n,L)
	\ge \min_{P_{XY} \colon P_X=Q}
	\min_{\tilde{P}_{XY} \colon \tilde{P}_X=Q,\ \tilde{P}_Y=P_Y}\\
	&\hspace{2em} D(P_{XY}\| Q \times W)
	+
	\left[ I_{\tilde{P}}(X \colon Y) - R \right]_+\\
	&\hspace{2em}+ L \left[ g(P_{XY}) - g(\tilde{P}_{XY}) - \left[ R - I_{\tilde{P}}(X \colon Y) \right]_+ \right]_+.
\end{align*}

\subsection{Proof of Theorem~\ref{thm:integral}} \label{subsec:appendix-proof-thm-integral}

\section{Integral exponent}

For convenience, we collect the following results as lemmas.

\begin{lemma}\label{lemma:chernoff}
	Let $X \sim \Bin\left(m,p\right)$.
	
	(a)~If $p \le r \le 1$, then
	\begin{align} 
		e^{-m \left( d\left( r \| p \right) + \epsilon(m) \right)}
		\le \Pbb\left( X \ge rm \right)
		\le e^{-md\left( r\|p \right)}, \label{eq:lemma-binomial-tail-a1}
	\end{align}
	where $\epsilon(m) \to 0$, as $m\to\infty$.
	
	(b)~If $0 \le r \le p$, then
	\begin{align} 
		\frac{1}{2}
		\le \Pbb\left( X \ge rm \right)
		\le 1. \label{eq:lemma-binomial-tail-b2}
	\end{align}

	(c)~If $r < 0$, then $\Pbb\left( X \ge rm \right) = 1$ and, if $r>1$, then $\Pbb\left( X \ge rm \right) = 0$.
\end{lemma}
\begin{IEEEproof}
	Part~(a) is an application of Chernoff's bound. For part~(b), it suffices to note that, in that regime, $\Pbb\left( X \ge rm \right) \ge \Pbb\left( X \ge \E[X] \right)$. Part~(c) is immediate from the support of~$X$.
\end{IEEEproof}

\begin{lemma}[{\hspace{0.1em}\cite[pp.~166--167]{merhav2010}}] \label{lemma:bound-binary-divergence}
	Let $r,p \in \left[0,1\right]$; then
	\begin{equation}
		d(r\|p) > r \left( \log\frac{r}{p} -1 \right).
	\end{equation}
\end{lemma}

We are going to split the proof in two cases. Let $X \sim \Bin\left( e^{nA_n}, e^{-nB_n} \right)$, and denote, for each $n\in\N$,
\begin{equation*}
	I(n) \coloneqq
	\int_0^{\infty} e^{-n\bL\theta} \cdot \Pbb\left( N_n \ge e^{n\left( C_n-\theta \right)} \right)\d\theta.
\end{equation*}

\textit{Case $A \le B$:} in this case, for $n$ large enough, $A_n \le B_n$. For each $n\in\N$, $\theta > C_n \iff e^{n\left( C_n-\theta \right)} < 1$. So we split each integral
\begin{align}
	I(n)
	&= \underbrace{\int_0^{\left[C_n\right]_+} e^{-n\bL\theta} \cdot \Pbb\left( N_n \ge e^{n(C_n-\theta)} \right)\d\theta}_{I_1(n)}\nonumber\\
	&\quad+ \underbrace{\int_{\left[C_n\right]_+}^{\infty} e^{-n\bL\theta} \cdot \Pbb\left( N_n \ge e^{n(C_n-\theta)} \right)\d\theta}_{I_2(n)}.
\end{align}

\textit{(Term $I_1$).} For the first term, we trivially have
\begin{equation} \label{eq:I1-lower-bound}
	I_1(n) \ge 0, \quad n\in\N.
\end{equation}
We are going to show that we have
\begin{equation} \label{eq:I1-upper-bound}
	I_1(n) \dotle e^{-n\left(B-A+\bL[C]_+\right)}.
\end{equation}
If $C < 0$, then, for large enough $n$, $C_n < 0$, so $I_1(n) \to 0$ and we are done. If $C=0$, then $I_1(n) \le [C_n]_+ \cdot 1 \to 0$, and we are done.
So we consider the case $C > 0$, and apply Lemma~\ref{lemma:chernoff} to $I_1(n)$ with $m_n = e^{nA_n}$, $p_n = e^{-nB_n}$ and $r_n = e^{-n(A_n+\theta-C_n)}$. We note that $p_n \le r_n \le 1 \iff C_n-A_n \le \theta \le C_n - A_n + B_n$, and in that regime we are in the case of~\eqref{eq:lemma-binomial-tail-a1}. Moreover, note that, for $\theta < C_n - A_n$, $r_n>1$ and $\Pbb\left(N_n \ge e^{-n(C_n-\theta)}\right)=0$, so we have
\begin{align*}
	I_1(n)
	&= \int_{0}^{[C_n]_+} e^{-n\bL\theta} \cdot \Pbb\left( N_n \ge e^{n(C_n-\theta)} \right) \d\theta\\
	&= \int_{[C_n-A_n]_+}^{[C_n]_+} e^{-n\bL\theta} \cdot \Pbb\left( N_n \ge e^{n(C_n-\theta)} \right) \d\theta\\
	&\le \int_{[C_n-A_n]_+}^{[C_n]_+} e^{-n\bL\theta}\\
	&\hspace{1em}\times \exp\left( -e^{nA_n} d\left( e^{-n(A_n + C_n - \theta)} \middle\| e^{-nB_n} \right) \right) \d\theta\\
	&\le \int_{[C_n-A_n]_+}^{C_n} e^{-n\bL\theta} \\
	&\hspace{0em} \times \exp\left( -n\left(C_n + B_n - A_n - \theta - \frac{1}{n}\right)e^{n(C_n-\theta)} \right)\d\theta\\
	&= \frac{e^{-n\bL C_n}}{n} \int_{0}^{n \min\left\{C_n,A_n\right\}}\\
	&\hspace{3em} \exp\left( -n\left(B_n-A_n\right)e^{y} -ye^{y} + \bL y + e^y \right)\d y,
\end{align*}
where in the last inequality we applied Lemma~\ref{lemma:bound-binary-divergence}, and in the last equality, we used the change of variables $y = n(C_n-\theta)$.
Denote $F_n(y) \coloneqq (B_n-A_n)e^y$ and $\psi(y) \coloneqq -ye^{y} + \bL y + e^y$. Note that $F_n(y) \ge B_n-A_n > 0$, for large $n$, and\footnote{
	We have $\psi''(y) = -e^y(1+y) < 0$, showing the function is concave. We can find the maximum by setting $\psi'(y) = \bL - ye^y = 0$, which solves to $y = W(\bL)$.
} $\psi(y) \le W(\bL)$. We then have, for large enough~$n$,
\begin{align*}
	I_1(n)
	&\le \frac{e^{-n \bL C_n}}{n} \left( n \min\left\{C_n,A_n\right\} \right) e^{-n(B_n-A_n)} e^{W(\bL)}\\
	&\doteq e^{-n\left( B-A + LC \right)}.
\end{align*}

\textit{(Term $I_2$).} For each $n\in\N$, in the regime $\theta > C_n \iff e^{n\left( C_n-\theta \right)} < 1$, we have
\begin{align*}
	\Pbb\left( N_n \ge e^{n\left( C_n-\theta \right)} \right)
	&\ge \Pbb\left( N_n = 1 \right)\\
	&= \binom{e^{nA_n}}{1} e^{-nB_n} \left( 1 - e^{-nB_n} \right)^{e^{nA_n}-1}\\
	&= e^{-n(B_n-A_n)} e^{n\Delta(n)},
\end{align*}
where
$\Delta(n) \coloneqq \frac{e^{nA_n}-1}{n} \log \left( 1 - e^{-nB_n} \right) \to 0$, for $A \le B$. So we have the lower bound
\begin{align} \label{eq:I2-lower-bound}
	I_2(n)
	&\ge \int_{\left[ C_n \right]_+}^{\infty} e^{-n\bL\theta} e^{-n(B_n-A_n)} e^{n\Delta(n)}\d\theta \nonumber\\
	&= e^{n\left( A_n-B_n+\Delta(n) \right)} \frac{e^{-n\bL\left[C_n\right]_+}}{n\bL}
	\doteq e^{-n\left( B-A+\bL\left[C\right]_+ \right)}.
\end{align}
In that regime, we also have, by Markov's inequality, for each $n\in\N$,
\begin{align*}
	\Pbb\left( N_n \ge e^{n(C_n-\theta)} \right)
	= \Pbb\left( N_n \ge 1 \right) \le \frac{\E\left[ N_n \right]}{1} = e^{n(A_n-B_n)},
\end{align*}
so we can also upper bound
\begin{align} \label{eq:I2-upper-bound}
	I_2(n)
	&\le e^{n(A_n-B_n)} \int_{\left[C_n\right]_+}^{\infty} e^{-n\bL\theta} \d\theta \nonumber\\
	&= e^{n(A_n-B_n)} \frac{e^{-n\bL\left[ C_n \right]_+}}{n\bL} \nonumber\\
	&\doteq e^{-n\left( B-A+\bL[C]_+ \right)}.
\end{align}

Combining \eqref{eq:I1-lower-bound}, \eqref{eq:I1-upper-bound}, \eqref{eq:I2-lower-bound} and \eqref{eq:I2-upper-bound}, and noting that $A \le B$, we find that
\begin{align}
	I(n)
	\doteq e^{-n\left(B-A+\bL[C]_+\right)}. \label{eq:thm-integral-case-1}
\end{align}

\textit{Case $A>B$:} in this case, for sufficiently large $n$, we have $A_n > B_n$. Note that, for each $n \in \N$, $\theta \ge C_n-A_n+B_n \iff e^{n(C_n-\theta)} \le e^{n(A_n-B_n)} = \E\left[ N_n \right]$. We split each integral in
\begin{align}
	I(n)
	&= \underbrace{\int_0^{\left[C_n-A_n+B_n\right]_+} e^{-n\bL\theta} \cdot \Pbb\left( N_n \ge e^{n(C_n-\theta)} \right)\d\theta}_{I_3(n)} \nonumber\\
	&\quad+ \underbrace{\int_{\left[C_n-A_n+B_n\right]_+}^{\infty} e^{-n\bL\theta} \cdot \Pbb\left( N_n \ge e^{n(C_n-\theta)} \right)\d\theta}_{I_4(n)}.
\end{align}

\textit{(Term $I_3$).} We trivially have
\begin{equation} \label{eq:I3-lower-bound}
	I_3(n) \ge 0, \quad n\in\N.
\end{equation}
We are going to show that
\begin{equation} \label{eq:I3-upper-bound}
	I_3(n) \dotle e^{-n \bL \left[ C-A+B \right]_+}.
\end{equation}
If $C-A+B < 0$, then, for $n$ large enough, $C_n-A_n+B_n < 0$, so that $I_3(n) \to 0$, and we are done.
If $C-A+B = 0$, then $I_3(n) \le \left[ C_n - A_n - B_n \right]_+ \cdot 1 \to [C-A-B]_+ = 0$, and we are done as well.	
So consider $C-A+B > 0$. Further split
\begin{align*}
	I_3(n)
	&= \underbrace{\int_{0}^{C_n-A_n+B_n-\frac{1}{n}} e^{-n\bL\theta} \cdot \Pbb\left( N_n \ge e^{n(C_n-\theta)} \right) \d\theta}_{I_{3,a}(n)}\\
	&+ \underbrace{\int_{C_n-A_n+B_n-\frac{1}{n}}^{C_n-A_n+B_n} e^{-n\bL\theta} \cdot \Pbb\left( N_n \ge e^{n(C_n-\theta)} \right) \d\theta}_{I_{3,b}(n)}.
\end{align*}
For the first term, apply Lemma~\ref{lemma:chernoff} with $m_n=e^{nA_n}$, $p_n = e^{-nB_n}$ and $r_n = e^{-n(A_n+\theta-C_n)}$. Again, for $p_n \le r_n \le 1 \iff C_n-A_n \le \theta \le C_n - A_n + B_n$, we are in the case of~\eqref{eq:lemma-binomial-tail-a1}, and for $\theta < C_n-A_n$, we have $r_n>1$ and $\Pbb\left( N_n \ge e^{n(C_n-\theta)} \right) = 0$. Thus,
\begin{align*}
	I_{3,a}(n)
	&= \int_{0}^{C_n-A_n+B_n-\frac{1}{n}} e^{-n\bL\theta} \cdot \Pbb\left( N_n \ge e^{n(C_n-\theta)} \right) \d\theta\\
	&= \int_{\left[C_n-A_n\right]_+}^{C_n-A_n+B_n-\frac{1}{n}} e^{-n\bL\theta} \cdot \Pbb\left( N_n \ge e^{n(C_n-\theta)} \right) \d\theta\\
	&\le \int_{\left[C_n-A_n\right]_+}^{C_n-A_n+B_n-\frac{1}{n}} e^{-n\bL\theta} \\
	&\hspace{1em}\times \exp\left( -e^{nA_n} d\left( e^{-n(A_n+\theta-C_n)} \middle\| e^{-nB_n} \right) \right) \d\theta\\
	&\le \int_{\left[C_n-A_n\right]_+}^{C_n-A_n+B_n-\frac{1}{n}} e^{-n\bL\theta}\\
	&\times \exp\left( - n e^{n(C_n-\theta)} \left( C_n-A_n+B_n-\theta - \frac{1}{n} \right) \right) \d\theta\\
	&= e^{-n\bL\left( C_n-A_n+B_n-\frac{1}{n} \right)}
	\int_{0}^{B_n + \min\left\{ C_n-A_n,\, 0 \right\}-\frac{1}{n}}\\
	&\hspace{4em} \exp \left( -nu \left( e^{n\left( u+A_n-B_n+\frac{1}{n} \right)} - \bL \right) \right) \d u,
\end{align*}
where in the last inequality we applied Lemma~\ref{lemma:bound-binary-divergence}, and in the last equality we used the change of variables $u=C_n-A_n+B_n-\frac{1}{n}-\theta$.
For large $n$, $B_n+\min\left\{ C_n-A_n,\, 0 \right\} - \frac{1}{n}> 0$, so that $u>0$.
Also, for large $n$, $A_n-B_n+\frac{1}{n} > 0$.
So we eventually have $e^{n\left( u+A_n-B_n+\frac{1}{n} \right)} > \bL$, since $\bL$ is fixed.
In that regime, we have $\exp\left( -nu \left( e^{n\left( u+A_n-B_n+\frac{1}{n} \right)} - \bL \right) \right) < 1$, and thus
\begin{align*}
	I_{3,a}(n)
	&\le e^{-n \bL \left( C_n-A_n+B_n-\frac{1}{n} \right)}\\
	&\hspace{2em} \times \left( B_n + \min\left\{ C_n-A_n,\, 0 \right\}-\frac{1}{n} \right)\\
	&\doteq e^{-n \bL \left( C-A+B \right)}\\
	&= e^{-n \bL \left[ C-A+B \right]_+}.
\end{align*}
For the second term, we immediately have
\begin{align*}
	I_{3,b}(n)
	&\le \int_{C_n-A_n+B_n-\frac{1}{n}}^{C_n-A_n+B_n} e^{-n\bL\theta} \d\theta\\
	&\le \frac{1}{n} e^{-n\bL\left( C_n-A_n+B_n-\frac{1}{n} \right)}\\
	&\doteq e^{-n\bL\left( C-A+B \right)}\\
	&= e^{-n \bL\left[ C-A+B \right]_+}.
\end{align*}

\textit{(Term $I_4$).} For each $n\in\N$, when $\theta \ge C_n-A_n+B_n \iff e^{n(C_n-\theta)} \le e^{n(A_n-B_n)} = \E\left[ N_n \right]$, we have
\begin{align*}
	\Pbb\left( N_n \ge e^{n(C_n-\theta)} \right) \ge \frac{1}{2},
\end{align*}
so that
\begin{align} \label{eq:I4-lower-bound}
	I_4(n)
	&\ge \int_{\left[ C_n-A_n+B_n \right]_+}^{\infty} e^{-n\bL\theta} \left(\frac{1}{2}\right) \d\theta \nonumber\\
	&= \frac{1}{2} \frac{e^{-n\bL\left[ C_n-A_n+B_n \right]_+}}{n\bL} \nonumber\\
	&\doteq e^{-n\bL\left[ C-A+B \right]_+}.
\end{align}
We also have
\begin{align} \label{eq:I4-upper-bound}
	I_4(n)
	&\le \int_{[C_n-A_n+B_n]_+}^{\infty} e^{-n\bL\theta}\nonumber\\
	&= \frac{e^{-n\bL[C_n-A_n+B_n]_+}}{n\bL} \nonumber\\
	&\doteq e^{-n\bL[C-A+B]_+}.
\end{align}

Combining \eqref{eq:I3-lower-bound}, \eqref{eq:I3-upper-bound}, \eqref{eq:I4-lower-bound} and \eqref{eq:I4-upper-bound}, and noting that $A>B$, we have
\begin{equation} \label{eq:thm-integral-case-2}
	I(n)
	\doteq e^{-n\bL \left[ C-A+B \right]_+}.
\end{equation}

\textit{Conclusion:}
We can combine results \eqref{eq:thm-integral-case-1}, for $A\le B$, and \eqref{eq:thm-integral-case-2}, for $A>B$. In any case, we have
\begin{equation*}
	I(n) \doteq
	e^{-n 
		\left(
		\left[ B-A \right]_+  + \bL \left[ C - \left[ A - B \right]_+ \right]_+
		\right)
	}.
\end{equation*}
In the particular case $\bL=1$, we can further simplify. If $A \le B$, then
\begin{align*}
	I(n)
	\doteq e^{-n\left(B-A+[C]_+\right)}
	= e^{-n\left[B-A+[C]_+\right]_+}.
\end{align*}
If $A > B$, then
\begin{equation*}
	I(n)
	\doteq e^{-n \left[ C-A+B \right]_+}
	= e^{-n \left[ B-A+[C]_+ \right]_+},
\end{equation*}
where we used the identity $\left[ a-b \right]_+ = \left[ [a]_+ - b \right]_+$ with $a=C$ and $b=A-B>0$.

\end{document}